\documentclass[prd,aps,nofootinbib,twocolumn,preprintnumbers,superscriptaddress]{revtex4-2}
\usepackage[utf8]{inputenc}
\usepackage[colorlinks=true,citecolor=blue,linkcolor=blue]{hyperref}
\usepackage[normalem]{ulem}
\usepackage{amsmath,amssymb, mathrsfs}
\usepackage{epsfig}
\usepackage{graphicx}               % Standard graphics package
\usepackage{url}
\usepackage{color}
\usepackage{slashed}
\usepackage{multirow}
\usepackage{placeins}
\usepackage[dvipsnames]{xcolor}
\usepackage{epstopdf}
\usepackage{soul}
\usepackage{tikz}
\usepackage[capitalise, english]{cleveref}
\usepackage{siunitx}
\usepackage{xspace}
\usepackage{booktabs}
\usetikzlibrary{trees}
\usetikzlibrary{decorations.pathmorphing}
\usetikzlibrary{decorations.markings}
\usepackage{comment}

\newcommand\myshade{80}
\colorlet{mylinkcolor}{ForestGreen}
\colorlet{mycitecolor}{Red}
\colorlet{myurlcolor}{violet}

\hypersetup{
  linkcolor  = mylinkcolor!\myshade!black,
  citecolor  = mycitecolor!\myshade!black,
  urlcolor   = myurlcolor!\myshade!black,
  colorlinks = true
}

\definecolor{jblue}{RGB}{20,50,100}
\definecolor{npurple}{RGB} {153, 51, 204}
\definecolor{wred}{RGB}{217,0,56}
\definecolor{white}{RGB}{255,255,255}

\definecolor{korange}{RGB}{235, 80,  43}
\definecolor{korange2}{RGB}{245, 100,  63}
\definecolor{kyelloworange}{RGB}{255, 210,  110}
\definecolor{kyelloworange2}{RGB}{240, 170,  90}
\definecolor{kred}{RGB}{204,  102, 153}
\definecolor{kpurple}{RGB}{153,  61, 190}
\definecolor{kpurplelight}{RGB}{213,  161, 230}

 \definecolor{tobycolour}{rgb}{.5,.0,.5}

\DeclareSIUnit\year{yr}
\DeclareSIUnit\pc{pc}
\DeclareSIUnit\ergs{ergs}
\DeclareSIUnit\msun{\ensuremath{M_\odot}}
\allowdisplaybreaks

\renewcommand{\vec}[1]{{\mathbf{#1}}}

\makeatletter
\providecommand*{\diff}%
  {\@ifnextchar^{\DIfF}{\DIfF^{}}}
\def\DIfF^#1{%
  \mathop{\mathrm{\mathstrut d}}%
    \nolimits^{#1}\gobblespace}
\def\gobblespace{%
  \futurelet\diffarg\opspace}
\def\opspace{%
  \let\DiffSpace\!%
  \ifx\diffarg(%
    \let\DiffSpace\relax
  \else
    \ifx\diffarg[%
      \let\DiffSpace\relax
    \else
        \ifx\diffarg\{%
        \let\DiffSpace\relax
      \fi\fi\fi\DiffSpace}      
      
\usepackage{tikz,xcolor,hyperref}

\definecolor{lime}{HTML}{A6CE39}
\DeclareRobustCommand{\orcidicon}{\hspace{-1mm}
	\begin{tikzpicture}
	\draw[lime, fill=lime] (0,0) 
	circle [radius=0.16] 
	node[white] {{\fontfamily{qag}\selectfont \tiny \,ID}};
	\draw[white, fill=white] (-0.0525,0.095) 
	circle [radius=0.007];
	\end{tikzpicture}
	\hspace{-3mm}
}

\foreach \x in {A, ..., Z}{\expandafter\xdef\csname orcid\x\endcsname{\noexpand\href{https://orcid.org/\csname orcidauthor\x\endcsname}
			{\noexpand\orcidicon}}
}

\begin{document}

%=============================================================================
\title{Dive deeper with SUBMARINE: SUB-Mev dArk matter diRect detectIon using bilayer grapheNE}

\author{Rinchen Sherpa\orcidA{}}
\email{rinchens@iisc.ac.in}
\affiliation{Centre for High Energy Physics, Indian Institute of Science, C.\,V.\,Raman Avenue, Bengaluru-560012, India}

\author{Anuvab Sarkar\orcidB{}} 
\email{sarkar.176@osu.edu}
\affiliation{Centre for High Energy Physics, Indian Institute of Science, C.\,V.\,Raman Avenue, Bengaluru-560012, India}
\affiliation{Department of Physics, Ohio State University, 191 West Woodruff Avenue, Columbus, OH 43210, USA}

\author{Tarak Nath Maity\orcidC{}} 
\email{tarak.maity.physics@gmail.com}
\affiliation{School of Physics, The University of Sydney and ARC Centre of Excellence for Dark Matter Particle Physics, NSW 2006, Australia}

\author{Paramita Dutta\orcidD{}} 
\email{paramita@prl.res.in}
\affiliation{Theoretical Physics Division, Physical Research Laboratory, Navrangpura, Ahmedabad-380009, India}

\author{Ranjan Laha\orcidE{}} 
\email{ranjanlaha@iisc.ac.in}
\affiliation{Centre for High Energy Physics, Indian Institute of Science, C.\,V.\,Raman Avenue, Bengaluru-560012, India}

\author{Anirban Das\orcidF{}} 
\email{anirbandas.21@protonmail.com}
\affiliation{Theory Division, Saha Institute of Nuclear Physics, 1/AF, Bidhannagar, Kolkata-700064, India}
\affiliation{Homi Bhabha National Institute, Training School Complex, Anushaktinagar, Mumbai-400094, India}
\date{\today}

%%%%%%%%%%%%%%%%%%%%%%%%%%%%%%%%%%%%%%%%%%%%%%%%%%%%%%%%%%%%%%%%%%%

\begin{abstract}
Novel target materials with anisotropic response will play a key role in detecting low-mass dark matter in upcoming experiments. Bilayer graphene is one such material that has been proposed for the detection of sub-MeV mass dark matter particles via electronic excitations. In this work, we calculate scattering rate via a massive mediator in bilayer graphene. With an exposure as small as $\sim$ 0.5 mg-year, bilayer graphene can probe new regions of the parameter space. The anisotropic response function of bilayer graphene leads to a sidereal-day modulation in the scattering rate, depending on its orientation with respect to the Galactic dark matter wind. We find significant modulation in the scattering rate for sub-MeV mass dark matter, demonstrating bilayer graphene's promise for a future experiment. We hope that our work will motivate the community to investigate bilayer graphene as a novel target material, and that it may lead us to discover the particle nature of dark matter.
\end{abstract}

\maketitle

\section{Introduction}
\label{sec:introduction}A large number of astrophysical and cosmological observations hint towards the existence of dark matter (DM)\,\cite{Cirelli:2024ssz}. Although we observe its gravitational effects, we do not understand what it is at a fundamental level. Over the last few decades, numerous experiments have tried to look for the potential particle nature of DM\,\cite{PhysRevD.31.3059}. While making significant progress, these experiments have not found any unambiguous DM signal so far. The first generation direct detection experiments were motivated by weakly interacting massive particles (WIMPs) which have masses above a GeV\,\cite{DarkSide-50:2025lns,CRESST:2025tol,PandaX:2024qfu,LZ:2024zvo,XENON:2025vwd}. 
However, there are many well-motivated models that predict DM masses below a GeV, and in recent years, the DM direct detection community has started exploring this sub-GeV mass regime\,\cite{Boehm:2003hm, Gori:2025jzu, Berlin:2017ftj, Chang_2021, Zurek:2024qfm,PhysRevLett.116.221302,PhysRevLett.113.171301,Essig:2022dfa}.

For DM masses below GeV, many-body effects in the target material may start to become significant\,\cite{Kahn:2021ttr}. Unlike WIMPs, which scatter with a single nucleus, such low-mass DM particles can deposit energy by exciting collective modes in the target\,\cite{Hochberg:2025rjs,Taufertshofer:2025agm,Liang:2024lkk,Catena:2025sxz,Catena:2024rym,Catena:2021qsr,Catena:2019gfa,Hochberg_2021,Knapen:2021run}. This understanding has motivated searches and proposals for novel target materials utilizing various collective excitations to detect such DM particles. 
These include electronic excitations in Dirac materials and topological semimetals\,\cite{Geilhufe:2019ndy,Hochberg:2017wce,Huang:2023nms}, molecular excitations\,\cite{Essig:2019kfe}, organic compounds\,\cite{Blanco:2025sgv,Blanco:2021hlm,Blanco:2019lrf}, superfluid helium\,\cite{Hertel:2018aal,SPICE:2023tru,Maris:2017xvi}, Migdal effect\,\cite{Blanco:2022pkt,Liang:2022xbu,Berghaus:2022pbu}, single-phonon and multi-phonon excitations\,\cite{Coskuner:2021qxo,Knapen:2017ekk,Trickle:2019nya,Stratman:2024sng,Mitridate:2023izi},
magnon excitations\,\cite{Trickle:2019ovy,Esposito:2022bnu,Marocco:2025eqw, Berlin:2025uka}, pressure tunable targets\,\cite{Ashour:2024xfp}, and quantum devices\,\cite{Hochberg:2026mdl,Das:2022srn,Das:2024jdz}. 

Beyond theoretical proposals, several experiments like CRESST, SuperCDMS, NEWS-G, QROCODILE, TESSERACT, SENSEI and DAMIC-M have begun probing this mass range through DM–nucleus scattering and DM-electron scattering\,\cite{Crisler:2018gci,DAMIC-M:2023gxo,TESSERACT:2025tfw,Angloher:2025fzw,SuperCDMS:2020aus,NEWS-G:2024jms,Baudis:2025zyn}. More recently, readout sensors such as transition-edge sensors (TESs), superconducting nanowire single-photon detectors (SNSPDs), and kinetic inductance detectors (KIDs) have begun to constrain portions of the sub-GeV DM parameter space, with the detector elements themselves serving simultaneously as both the sensor and target\,\cite{Baudis:2025zyn,Schwemmbauer:2025evp, Gao:2024irf}. These technologies appear promising, with ongoing developments expected to further lower energy thresholds and improve sensitivity in future experiments. Many upcoming experiments, currently at their research and development phase, will also target sub-GeV DM parameter space\,\cite{Abbamonte:2025guf,DELight:2024bgv,SPICE:2023tru}. 
In addition to new experiments, results from existing experiments have been reinterpreted to constrain portions of sub-GeV DM parameter space\,\cite{Cheek:2025nul,Du:2024afd,Liang:2024xcx,Maity:2022exk,XENON:2019zpr,Essig:2012yx}. 
\begin{figure*}[!htbp]
\centering
\includegraphics[width =0.9 \columnwidth]{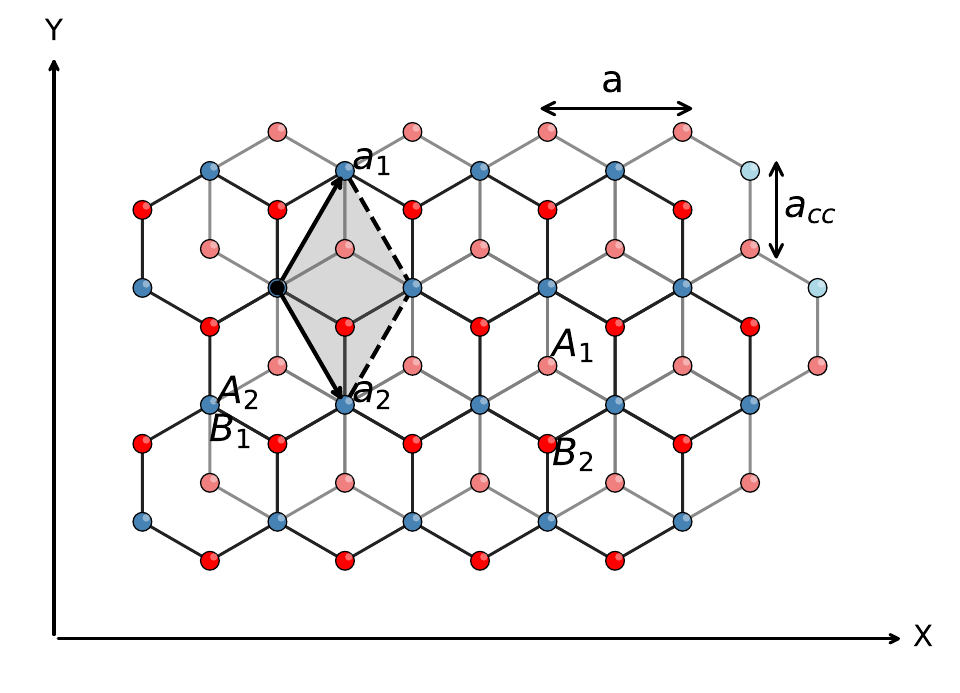}
\includegraphics[width =0.8 \columnwidth]{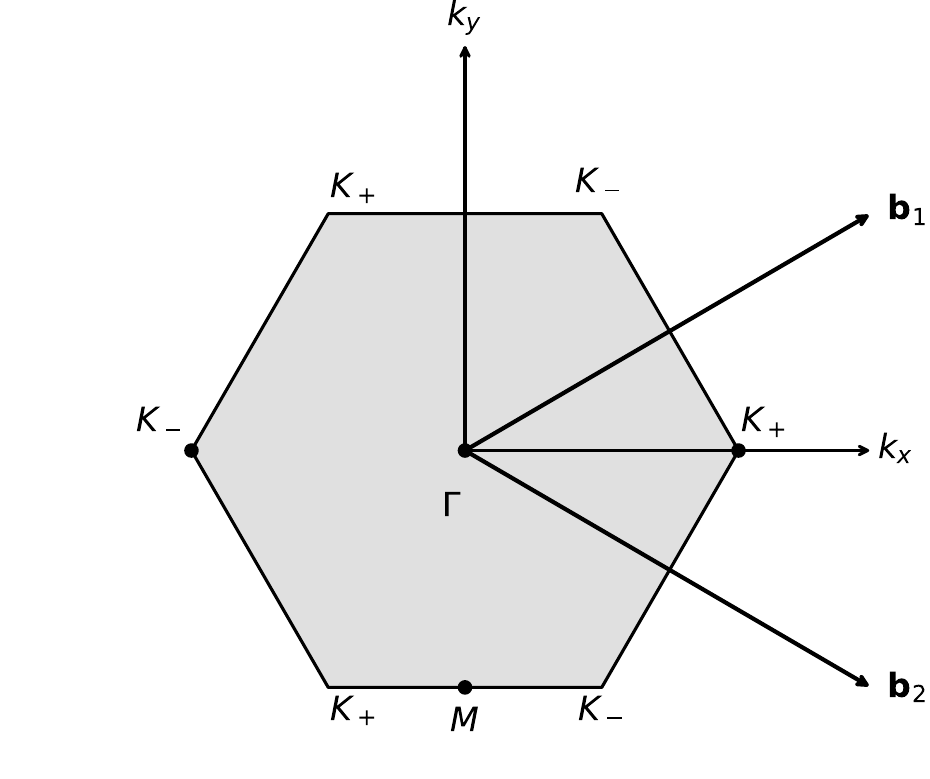}  
    \caption{Left: Lattice structure of bilayer graphene consisting of four inequivalent sites labelled by $A_1$, $B_1$, $A_2$ and $B_2$, (see also Fig.1 of Ref.\,\cite{Das:2023cbv}). The two layers are indicated by different opacities. The Grey shaded region indicates the unit cell with $\mathbf{a_1}$ and $\mathbf{a_2}$ as the lattice vectors and the side of each hexagon as $a_{\text{cc}}=1.42$\,\AA{} and $a=\sqrt{3}a_{\text{cc}}$. Right: The first Brillouin zone (BZ) with $\textbf{b}_1$ and $\textbf{b}_2$ as the reciprocal space lattice vectors, the centre point $\Gamma$, and other high symmetry points $M$, $K_+$, $K_-$ shown at the side and corners of the hexagon, respectively.} 
    \label{lattice structure}
\end{figure*}

Among various proposed materials, graphene is one such material that has been suggested for detecting sub-GeV DM via DM-electron scattering\,\cite{Wang:2015kya, Hochberg:2016ntt, Kim:2020coy, Catena:2023awl}. Graphene is an allotrope of carbon consisting of a single layer of atoms arranged in a two-dimensional ($2$D) honeycomb lattice. At low energies, the dispersion of electrons is linear, and thus the electronic excitations can be described as massless Dirac fermions\,\cite{Castro_Neto_2009}. Previous works utilizing monolayer graphene have mainly focused on electron ejection for detecting DM. Ref.\,\cite{Das:2023cbv} alternatively proposed the use of bilayer graphene (BLG), which consists of two stacked layers of monolayer graphene, for detecting sub-MeV mass DM particles via electronic excitations from the valence to conduction band. In BLG, a small tunable gap can be opened between the valence and conduction band by applying a gate voltage between two layers, which can be used as an energy threshold to look for excitations due to DM-electron scattering and isolate the signal from the  background\,\cite{Das:2023cbv,Oostinga_2007,doi:10.1126/science.1130681}. Moreover, the $2$D nature of BLG makes its response naturally anisotropic. Such anisotropy can give rise to sidereal modulation in the scattering rate. This is particularly useful for distinguishing the DM signal from various backgrounds. 

In this work, we build upon the work of Ref.\,\cite{Das:2023cbv} by performing the calculations for DM-electron scattering via a massive mediator. We also examine the effect of BLG orientation on the daily modulation of the scattering rate. Note that, for consistency, we perform all calculations for both massless and massive mediators. 

The rest of the paper is organized as follows. We discuss the structure of BLG and describe it by using the tight-binding framework in Sec.\,\ref{sec:Lattice structure}. The method for calculating the DM-electron scattering rate and the dynamic structure factor for BLG are illustrated in Secs.\,\ref{sec:scattering} and
\ref{sec: Structure factor}, respectively. The results for the scattering rate and projected sensitivity are analyzed in Sec.\,\ref{sec: projected sensitivity}. For the daily modulation of the scattering rate and its dependence on the orientation of the BLG, we refer to Sec.\,\ref {sec:Dailymod}. Finally, we summarize and conclude in Sec.\,\ref{sec:conclusions}.
Throughout the work, we consider natural units where $\hbar = c = \varepsilon_0 = 1$.

\begin{figure*}[!htbp]
\centering
\includegraphics[width = 1.9\columnwidth]{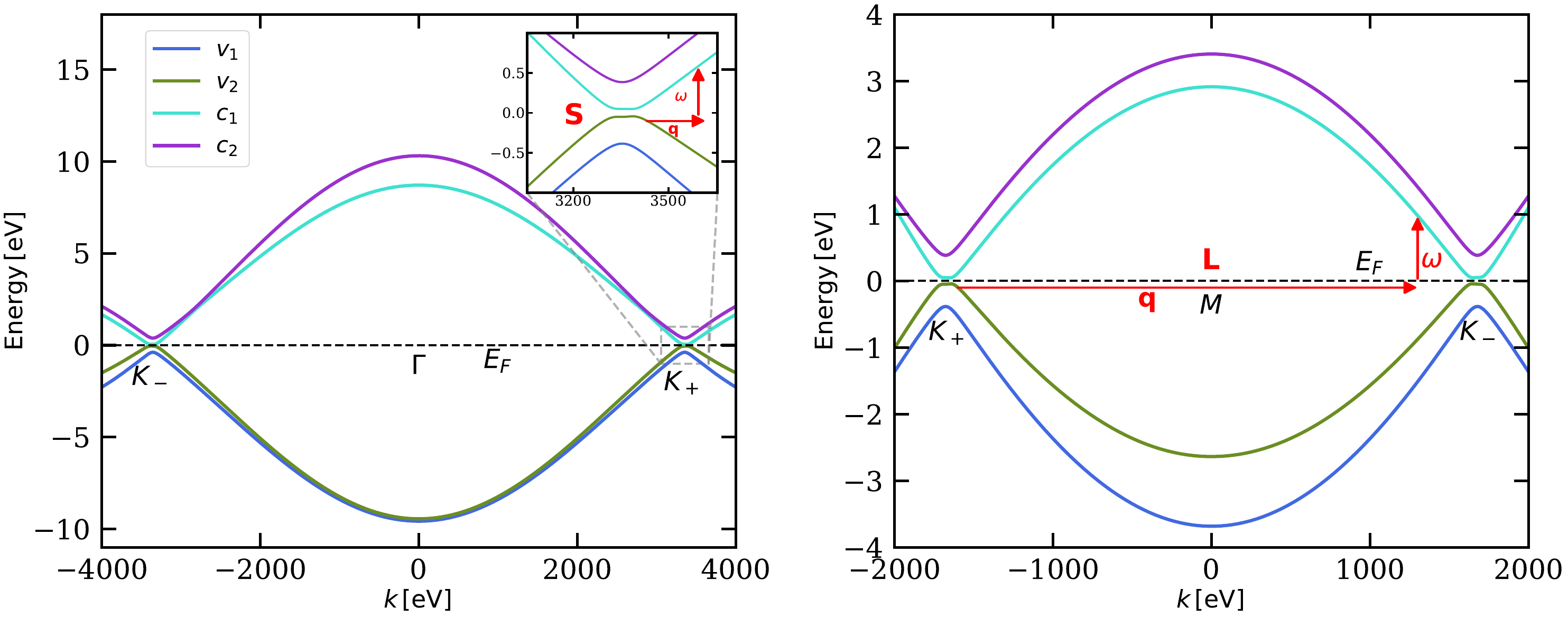}
\caption{Band structure of bilayer graphene along along $K_{-} \rightarrow \Gamma \rightarrow K_{+}$ path (left) and $K_{+} \rightarrow M \rightarrow K_{-}$ path (right). The two valence (conduction) bands are indicated by $v_1$ and $v_2$ ($c_1$ and $c_2$) and the Fermi energy level by $E_{\text{F}}$. An electron-hole excitation near the $K_+$ point, caused by DM-electron scattering, is shown in the left inset. Transitions  caused by small-energy and small-momentum transfers are indicated by the letter `S', whereas, the high momentum but low energy transfers are denoted by `L'.} 
\label{band structure}
\end{figure*}

\section{Bilayer Graphene}
\label{sec:Lattice structure}
The electronic configuration of carbon is $1s^22s^22p^2$. In graphene, the $2s$, $2p_x$, $2p_y$ orbitals combine to form three $sp^2$ hybridized orbitals. These hybridized $sp^2$ orbitals form $\sigma$-bonds with the neighboring atoms, whereas the $2p_z$ orbitals form $\pi$-bonds. The $\sigma$-bonds are aligned in a plane with $120^{\circ}$ angle between them, giving rise to a honeycomb lattice structure\,\cite{McCann_2011,Castro_Neto_2009}. BLG consists of two stacked graphene sheets. In this work, keeping in mind the stability of the BLG we focus on the Bernal stacking (AB) configuration, in which half of the carbon atoms in the top layer lie directly above atoms in the bottom layer, as illustrated in Fig.~\ref{lattice structure}\,\cite{J.-C.Charlier_1994}.
There are four inequivalent lattice sites in BLG labelled by $A_1$, $B_1$, in the first layer and $A_2$, $B_2$ in the second layer. The primitive lattice vectors of the BLG are,
\begin{equation}
\label{equ: primitive lattice vectors}
    \mathbf{a}_1 = \left( \frac{a}{2}, \frac{\sqrt{3}a}{2} \right), \quad
\mathbf{a}_2 = \left( \frac{a}{2}, -\frac{\sqrt{3}a}{2} \right),
\end{equation}
where $a = \sqrt{3} a_{\text{cc}}=2.46$\,\AA{} is the lattice constant and $a_{\text{cc}}=1.42$\,\AA{} is the side of each hexagon. The unit cell of the BLG consists of four carbon atoms and is shown by the shaded region in Fig.\,\ref{lattice structure}. The distance between the two layers is $d= 3.35$\,\AA. We define the BLG coordinate frame such that the $\mathrm{X-Y}$ axes lie in the plane of the BLG, while the $\mathrm{Z}$-axis is perpendicular to the BLG plane. The reciprocal lattice vectors of the BLG are given by,
\begin{equation}
\label{equ: reciprocal lattice vectors}
    \mathbf{b}_1 = \left( \frac{2\pi}{a}, \frac{2\pi}{\sqrt{3}a} \right), \quad
\mathbf{b}_2 = \left( \frac{2\pi}{a}, -\frac{2\pi}{\sqrt{3}a} \right).
\end{equation}
and are shown in the first Brillouin zone (BZ) in Fig.\,\ref{lattice structure}. 
The center point $\Gamma$, the corner points $K_{\pm}$, and the midpoints $M$ of the hexagonal edges are the high-symmetry points. To describe the BLG, we use a tight-binding model of the BLG, taking one $2p_z$ $\pi$ orbital per atomic site. 
The Hamiltonian for this tight-binding model is written as (see Appendix \ref{app: tight binding model} for details),
\begin{equation}
\label{equ: tight binding hamiltonian}
H = \begin{pmatrix}
\varepsilon_{A1} & -\gamma_0 f(k) & \gamma_4 f(k) & -\gamma_3 f^{*}(k) \\
-\gamma_0 f^{*}(k) & \varepsilon_{B1} & \gamma_1 &\gamma_4 f(k)\\
\gamma_4 f^*(k) & \gamma_1 & \varepsilon_{A2} & -\gamma_0 f(k) \\
-\gamma_3 f(k) & \gamma_4 f^*(k) & -\gamma_0 f^*(k) & \varepsilon_{B2}
\end{pmatrix}\,,
\end{equation}
where diagonal elements, i.e., $\varepsilon_{A1}$, $\varepsilon_{B1}$, $\varepsilon_{A2}$, $\varepsilon_{B2}$, are the onsite potential energies, $\gamma_{0}$, $\gamma_{1}$, $\gamma_{3}$ and $\gamma_{4}$ are the off-diagonal elements, termed the hopping integrals, and  
$f(k) = e^{i k_y a / \sqrt{3}} + 2 e^{-i k_y a / (2 \sqrt{3})} \cos(k_x a / 2)$. 
In the presence of an applied gate voltage $U$, $\varepsilon_{A1}=\varepsilon_{B1}=-U/2$ and $\varepsilon_{A2}=\varepsilon_{B2}=U/2$, and we use other parameters as $\gamma_{0}= 3.16$ eV, $\gamma_{1}= 0.381$ eV, $\gamma_{3} = 0.38$ eV, $\gamma_{4} = 0.14$ eV\,\cite{McCann_2011,Kuzmenko_2009}. By diagonalizing the matrix $H$ of Eq.~\eqref{equ: tight binding hamiltonian}, we obtain the energy dispersion, which is shown along $K_-\rightarrow \Gamma \rightarrow K_+$ and $K_+ \rightarrow M \rightarrow K_-$ paths in Fig.~\ref{band structure}. The valence and conduction bands are labeled as shown in Fig.~\ref{band structure}. The band gap near the K points is given by $E_g = |U| \gamma_1/\sqrt{\gamma_1^2 + U^2} \approx |U|$ for small values of $U$, and this gap $E_g\equiv E_{th}$ is used as the threshold energy of the detector. The Fermi energy lies in the band gap, shown by a dashed line. 
 
\section{Scattering Rate} 
\label{sec:scattering}
With the tight-binding model description, we calculate the scattering rate of the DM in BLG. We consider spin-independent DM-electron interactions where the mediator at leading order only couples to the electron density in the non-relativistic limit and the momentum space potential is\,\cite{Knapen:2021run,Boyd_2023}
\begin{equation}
\label{equ: momentum space potential}
V(q) = \frac{g_e g_\chi}{q^2+m_\phi^2}\,,  
\end{equation} 
where $q$ is the magnitude of the momentum transfer, $g_e$ and $g_\chi$ are the mediator couplings to electrons and DM, respectively, and $m_\phi$ is the scalar (or vector) mediator mass. The differential scattering rate of the local DM with electrons from a target material of density $\rho_T$ is (see Appendix \ref{app: derivation of scattering rate} for the derivation)\,\cite{Trickle_2020, Kahn:2021ttr}
\begin{equation}
\label{equ: differential scattering rate}
\frac{dR}{d\omega} = \frac{\rho_\chi}{\rho_T m_\chi} \int \frac{d^3 q}{(2\pi)^3} \, \frac{\pi \, \sigma_{\chi e}}{\mu_{\chi e}^2} \, F_{\text{DM}}^2(q) \, S(\vec{q}, \omega) \, g(\vec{q}, \omega, t),
\end{equation}
where $R$ is the total scattering rate, $\omega$ is the transferred energy, $\rho_\chi$ is the local DM density, $m_\chi$ is the DM mass, $\mu_{\chi e}$ is the DM-electron reduced mass, $\sigma_{\chi e}$ is the DM-electron scattering cross section at the reference momentum $q_0=\alpha m_e$, where $\alpha$ is the fine structure constant and $m_e$ is the mass of the electron and $F_{\rm DM}(q)$ is the DM form factor,
\begin{equation}
\label{equ: DM form factor and ref cross-section}
\sigma_{\chi e} \equiv \frac{\mu_{\chi e}^2}{\pi} \left( \frac{g_e g_\chi}{q_0^2 + m_\phi^2} \right)^2 , \quad 
F_{\text{DM}}(q) = \frac{q_0^2 + m_\phi^2}{q^2 + m_\phi^2}\,.
\end{equation}
The response of the target material is given by the dynamic structure factor $S(\textbf{q},\omega)$ which we derive in the next section for BLG. The kinematic function is
\begin{equation}
\label{equ: G - function definition}
g(\vec{q}, \omega, t) = \int d^3 \vec{v} \, f(\vec{v}) \, \delta\left(\omega - \left(\textbf{q.v} - \frac{q^2}{2 m_\chi}\right)\right)\,,
\end{equation} 
where $\vec{v}$ is the DM velocity, $f(\textbf{v})$ is the lab-frame DM velocity distribution, and the delta function is due to energy-momentum conservation, i.e,
\begin{equation}
\label{equ: energy-momentum conservation}
    \omega = \textbf{q.v} - \frac{q^2}{2 m_\chi}.
\end{equation}
For the commonly assumed Maxwell-Boltzmann velocity distribution of DM, $g(\vec{q}, \omega, t)$  has an analytic form (see Appendix\,\ref{app: Derivation velocity integral}),
\begin{align}
\label{equ: G - function analytic form}
g(\vec{q}, \omega, t) &= \frac{\pi v_0^2}{q N_0} \left[ 
\exp\left( -\frac{v_{-}(q,t)^2}{v_0^2} \right) 
- \exp\left( -\frac{v_{\text{esc}}^2}{v_0^2} \right) 
\right], 
\end{align}
with
\begin{align}
v_{-}(\vec{q},t) &= \min \left\{ v_{\text{esc}}, \frac{\omega}{q} + \frac{q}{2 m_\chi} + \hat{q} \cdot \vec{v}_{\text{lab}}(t) \right\},
\end{align}
and 
\begin{equation}
\label{equ: G - function No normalization constant}
N_0 = \pi^{3/2} v_0^3 \left[ \mathrm{erf}\left( \frac{v_{\text{esc}}}{v_0} \right) - \frac{2 v_{\text{esc}}}{\pi^{1/2} v_0} \exp\left( -\frac{v_{\text{esc}}^2}{v_0^2} \right) \right]
\end{equation} is the normalization constant for the Maxwell-Boltzmann distribution with velocity dispersion $v_0$ and truncated at the escape velocity $v_{\rm esc}$ at the Solar circle. We have used $v_0 = 230$ km/s and $v_{\rm esc} = 500 $ km/s for our calculations same as in Ref.\,\cite{Das:2023cbv}. Here, $\vec{v}_{\text{lab}}(t)$ is the lab/Earth velocity in the Galactic frame.

Note that, the Galactic DM velocity distribution might not be isotropic Maxwellian, since recent mergers can make the distribution quite different\,\cite{Evans:2018bqy,Maity:2022enp}. Ref.\cite{Folsom:2025lly} used a suite of 98 Milky-Way-like galaxies from the IllustrisTNG50 simulation and showed that uncertainties due to these are much less compared to the systematics of the direct detection experiment. In addition to the Maxwell-Boltzmann velocity distributions, we also show our results using the DM velocity distribution functions as suggested by Ref.\,\cite{Folsom:2025lly}.

\section{Dynamic structure factor } 
\label{sec: Structure factor}

The scattering rate of Eq.\,\eqref{equ: differential scattering rate} depends on the dynamic structure factor, $S(\mathbf{q},\omega)$, which is a target material-specific quantity. Generally, the dynamic structure factor is a measure of density fluctuations of wavevector $\bold{q}$ and frequency $\omega$ in a system. It is expressed as a dynamical correlation function of the density, as shown in Eq.\,\eqref{equ: structure factor appendix}
\begin{equation}
     S(\bold{q},\omega)=\frac{1}{V} \int_{-\infty}^{\infty} dt e^{i \omega t} \langle \hat{n} (\vec{q}, t) \hat{n} (-\vec{q}, 0) \rangle\,,
\end{equation}
where $V$ is the volume, $\hat{n}(\boldsymbol{q},t)$ is the electron density operator in the momentum space at time $t$ and $\langle\,\rangle$ denotes the expectation over a thermal ensemble of states at finite temperature. For scattering, it is more natural to interpret $S(\bold{q},\omega)$ as the spectral density of excited states at wavevector $\vec{q}$ and energy $\omega$ that can be generated from the initial ground state by the density operator (see Appendix~\ref{app: derivation of scattering rate})\,\cite{Giuliani_Vignale_2005,book,Phillips_2012}.

For a 2D material such as BLG, the electronic density is confined near the plane Z=0 (where Z = 0 is in between the two layers). We separate the spatial coordinates into in-plane and out-of-plane components, $\mathbf r=(\mathbf r_\parallel,r_z)$, and approximate the electron density as uniform within the interlayer spacing $d$. This is a valid approximation as the inverse of the typical momentum transfer $\sim\frac{1}{q}$ $>$ $d$ for sub-MeV DM. Under this approximation, the density operator in the real space, which is the inverse-Fourier transform of $\hat{n}(q,t)$ defined as $\hat{n}(\vec{r},t)=\int d^3r \, e^{i \vec{q}.\vec{r}} \hat{n}(\vec{q},t)$, can be written as
\begin{equation}
\hat n(\mathbf r,t)=\hat n_{2D}(\mathbf r_\parallel,t)\,
\frac{1}{d}\Theta(d/2-|r_z|),
\end{equation}
with $d=3.35\,\text{\AA}$ as the separation between the graphene layers and $\Theta$ is the Heaviside step function. Substituting this form into the definition of $\hat n(\mathbf q,t)$ yields
\begin{equation}
\hat n(\mathbf q,t)=\hat n_{2D}(\mathbf q_\parallel,t)
\left(\frac{\sin(q_z d/2)}{q_z d/2}\right),
\end{equation}
where the factor in parentheses arises from the finite thickness of the electron distribution. Using this result, the 3D structure factor can be expressed in terms of the 2D structure factor as
\begin{equation}
S(\mathbf q,\omega)=\frac{1}{d}\,
S_{2D}(\mathbf q_\parallel,\omega)
\left(\frac{\sin(q_z d/2)}{q_z d/2}\right)^2.
\end{equation}

Using the linear response theory, the change in the electron density due to a weak perturbation that couples to the electron density (which is due to spin-independent DM-electron interactions in our case) will be proportional to the perturbation and the proportionality constant is the density-density response function, namely, polarization function~\cite{Bruus,tong_kinetic_theory_2012,Phillips_2012},
\begin{eqnarray}
\chi_{2D}(\mathbf{q}_{\parallel}, \omega) = && -\frac{i}{A}
\int dt \, \Theta(t) e^{i\omega t} \nonumber \\
&&\left\langle
\left[
\hat{n}_{2D}(\mathbf{q}_{\parallel}, t),
\hat{n}_{2D}(-\mathbf{q}_{\parallel}, 0)
\right] \right\rangle 
\end{eqnarray}
(see, for e.g., chapter 6 of Ref.\,\cite{Bruus}). 
A consequence of this is the relation between the structure factor and density-density response function,
\begin{equation}
S_{2D}(\mathbf q_\parallel,\omega)=-\frac{2}{1-e^{-\beta\omega}}
\,\mathrm{Im}\,\chi_{2D}(\mathbf q_\parallel,\omega),
\end{equation}
which is the well-known fluctuation-dissipation theorem and relates the equilibrium fluctuations, $S_{2D}(\mathbf q_\parallel,\omega)$, in a system to dissipation, $\mathrm{Im}\,\chi_{2D}(\mathbf q_\parallel,\omega)$, in a system\,\cite{Phillips_2012,tong_kinetic_theory_2012,Giuliani_Vignale_2005}. Here, $\beta$ is the inverse temperature.
At zero temperature the 2D structure factor reduces to
\begin{equation}
S_{2D}(\mathbf q_\parallel,\omega)=-2\,\Theta(\omega)\,
\mathrm{Im}\,\chi_{2D}(\mathbf q_\parallel,\omega).
\end{equation}
It is convenient to introduce the dielectric function, which is the longitudinal component of the dielectric tensor,
\begin{equation}
\varepsilon(\mathbf q_\parallel,\omega)=(1+v_c(\mathbf q_\parallel)\chi_{2D}(\mathbf q_\parallel,\omega))^{-1},
~
v_c(\mathbf q_\parallel)=\frac{e^2}{2\varepsilon_r \mathbf q_\parallel},
\end{equation}
where $v_c(\mathbf q_\parallel)$ is the Coulomb interaction in 2D and
$\varepsilon_r$ is the background dielectric constant, which we assume to be vacuum. The dynamic structure factor can then be written in terms of the energy-loss function (ELF),
\begin{equation}
\mathcal{W}( \mathbf q_\parallel,\omega)=\mathrm{Im}\!\left[-\frac{1}{\varepsilon( \mathbf q_\parallel,\omega)}\right],
\end{equation}
as
\begin{equation}
S_{2D}(\mathbf q_\parallel,\omega)=
\frac{2\Theta(\omega)}{v_c(\mathbf q_\parallel)}\,\mathcal{W}(\mathbf q_\parallel,\omega).
\end{equation}
Thus, in this formalism, the scattering rate is related to the ELF, an experimentally measurable quantity, and naturally incorporates many-body effects such as collective excitations and in-medium screening of the DM-electron interaction\,\cite{Hochberg_2021,Knapen:2021run}.

Within the random phase approximation (RPA), which corresponds to summing of infinite series of electron-hole bubble diagrams in perturbation theory, the interacting polarization function is\,\cite{Giuliani_Vignale_2005,book,Phillips_2012},
\begin{align}
    \chi_{2D} &= \chi_{2D}^0 + \chi_{2D}^0v_c\chi_{2D}^0 + \chi_{2D}^0v_c\chi_{2D}^0v_c\chi_{2D}^0+ ... \nonumber \\
    &=\chi_{2D}^0(1+v_c\chi_{2D}^0 +v_c\chi_{2D}^0v_c\chi_{2D}^0+..) \nonumber \\
    &=
\frac{\chi^0_{2D}(\mathbf q_\parallel,\omega)}{1-v_c(\mathbf q_\parallel)\chi^0_{2D}(\mathbf q_\parallel,\omega)}=\frac{\chi^0_{2D}(\mathbf q_\parallel,\omega)}{\varepsilon(\mathbf q_\parallel,\omega)}
\end{align}
where $\chi^0_{2D}$ is the noninteracting polarization function. 

The non-interacting polarization function for BLG is calculated using the single particle states as\,\cite{Giuliani_Vignale_2005,Dressel_Gruner_2002},
\begin{eqnarray}
\label{equ: non-interacting polarization function integral}
\chi_{2D}^{0}(\mathbf{q}_\parallel, \omega) &= &g \sum_{n,m} \int \frac{d^2 k}{(2\pi)^2} 
\frac{f_{\mathbf{k},n} - f_{\mathbf{k}+\mathbf{q}_\parallel,m}}{\omega + \varepsilon_{\mathbf{k},n} - \varepsilon_{\mathbf{k}+\mathbf{q}_\parallel,m} + i\eta} \nonumber \\
&& \times \left| \langle \mathbf{k}, n | \mathbf{k} + \mathbf{q}_\parallel, m \rangle \right|^2,\,\,\,
\end{eqnarray}
where the integral is over the first BZ, $\varepsilon_{\mathbf{k},n}$, $| \mathbf{k}, n \rangle$  are the eigenenergy and eigenvector respectively, for the wave vector $\mathbf{k}$ and band index $n$ obtained by diagonalizing the tight-binding Hamiltonian of Eq.\,\eqref{equ: tight binding hamiltonian}. Here, $f_{\bold{k},n}=(1+e^{\beta (\varepsilon_{\mathbf{k},n} - \mu)})^{-1}$ is the Fermi-Dirac distribution where $\mu$ is the chemical potential. The factor $g = 2$ accounts for the spin degeneracy of electrons.  The parameter $\eta$ is a phenomenological broadening term proportional to the inverse lifetime (or decay width) of quasiparticles. The exact value of $\eta$ does not affect our results, and we set $\eta = 1$ meV (typical order of magnitude for BLG) for numerical calculations\,\cite{PhysRevLett.102.176804}. We consider zero temperature throughout our calculations.

\begin{figure*}[!htbp]
\centering
\includegraphics[width=0.8\textwidth]{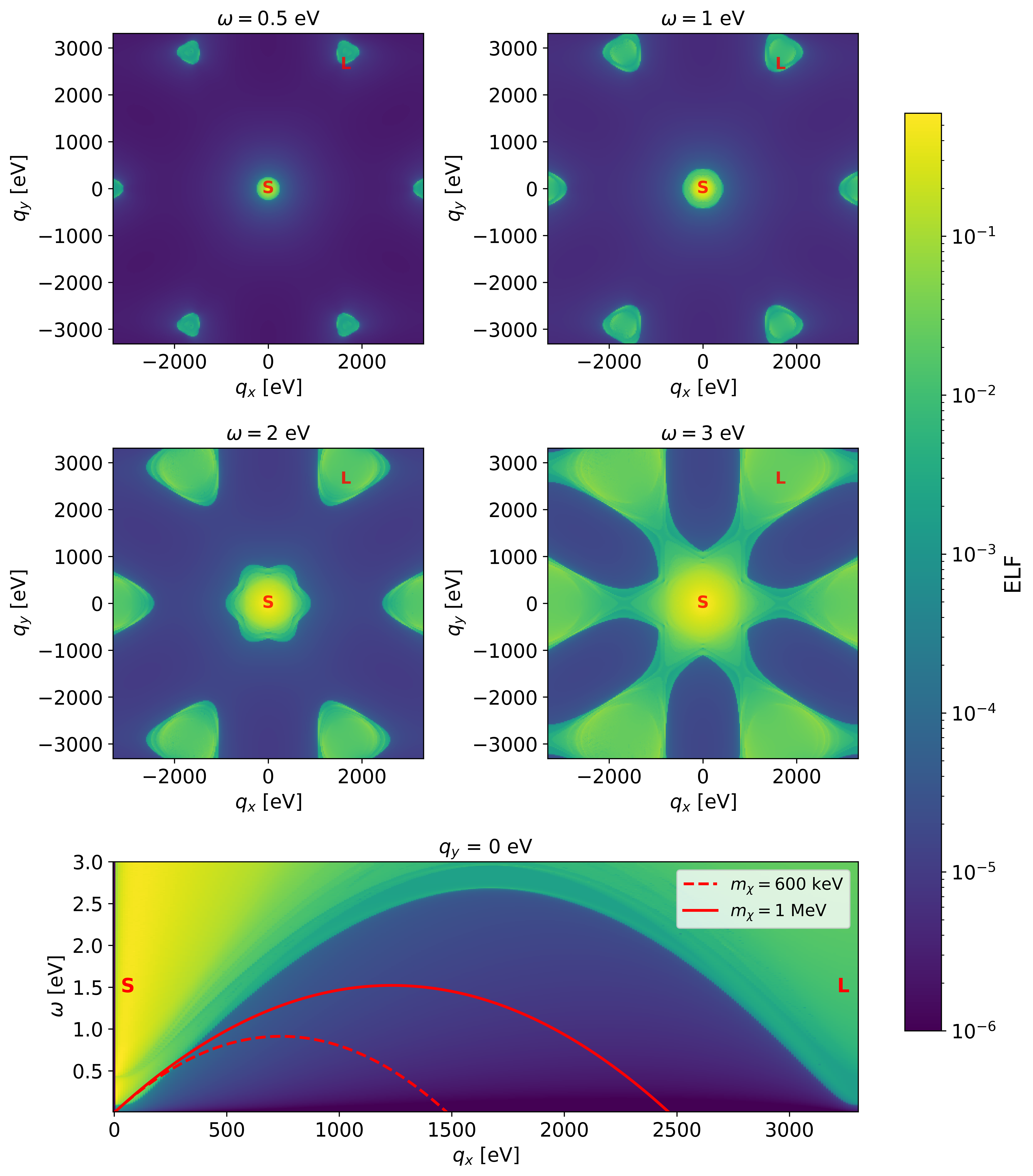}  
\caption{Color plot of BLG energy-loss function (ELF) as a function of in-plane momentum transfer ($q_x$ and $q_y$) (top four), for fixed energy transfer $\omega=0.5$ eV (top left), $\omega=1$ eV (top right), $\omega= 2$ eV (middle left) and $\omega= 3$ eV (middle right) with energy band gap of 100 meV.   Colour plot of BLG as a function of energy ($\omega$) and momentum transfer in the X-direction ($q_x$) (bottom). The solid and dashed red lines indicate the kinematically allowed boundaries for DM masses $m_\chi$ = 1 MeV and $m_\chi$ = 600 keV, respectively, obtained using the maximum possible velocity (740 km/s). Bright regions indicate the electron-hole continuum. The letters `S' and `L' correspond to the dominant kinds of transitions that contribute to the ELF in that region. } 
    \label{energy loss function}
\end{figure*}

\begin{figure*}[!htbp]
    \centering
    \includegraphics[width = 2\columnwidth]{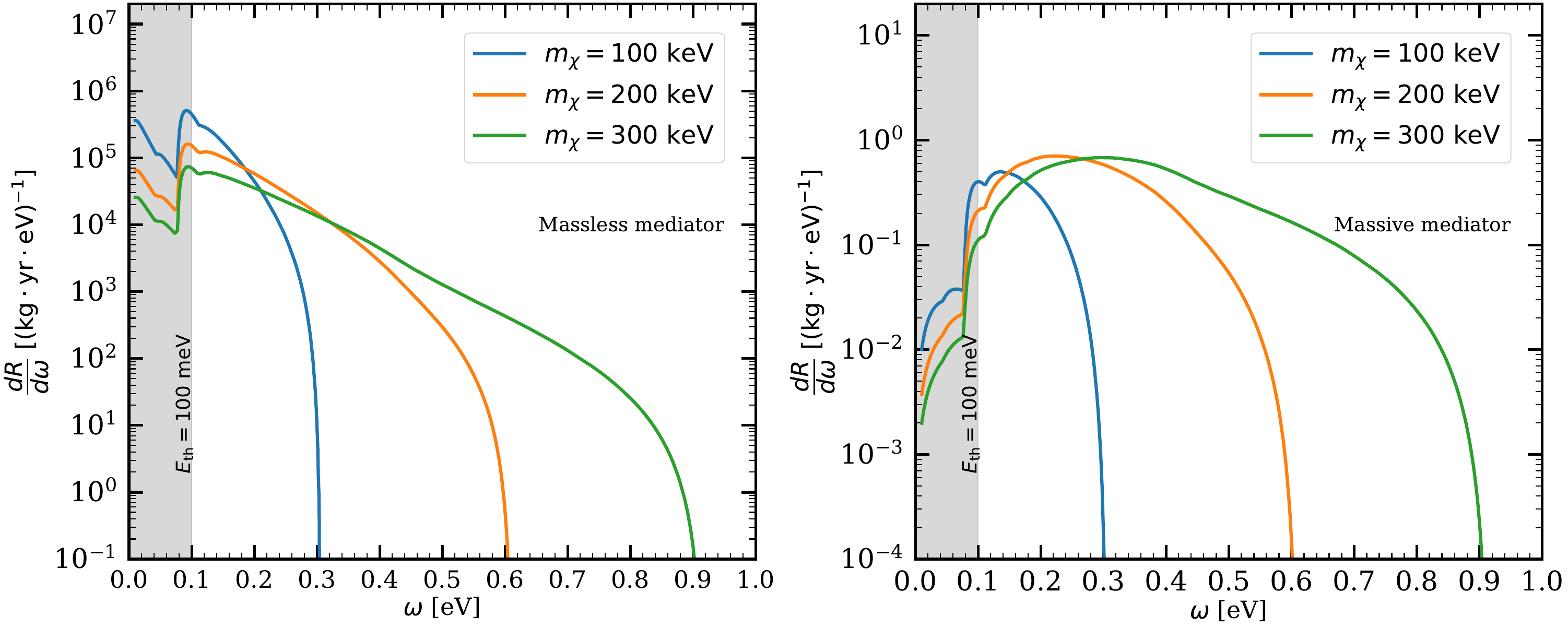} 
    \caption{Differential DM-electron scattering rate via massless (left) and massive mediator (right) in BLG at $t$ = 1 hr, with band gap $U$=100 meV and $\sigma_{\chi e}= 10^{-38}$ cm$^2$, for  DM masses $m_\chi$= 100 keV (blue), $m_\chi$= 200 keV (orange) and $m_\chi$= 300 keV (green).}.   
    \label{fig: differential scattering rate}
\end{figure*}

\begin{figure*}[!htbp]
    \centering
    \includegraphics[width = 1.03 \columnwidth]{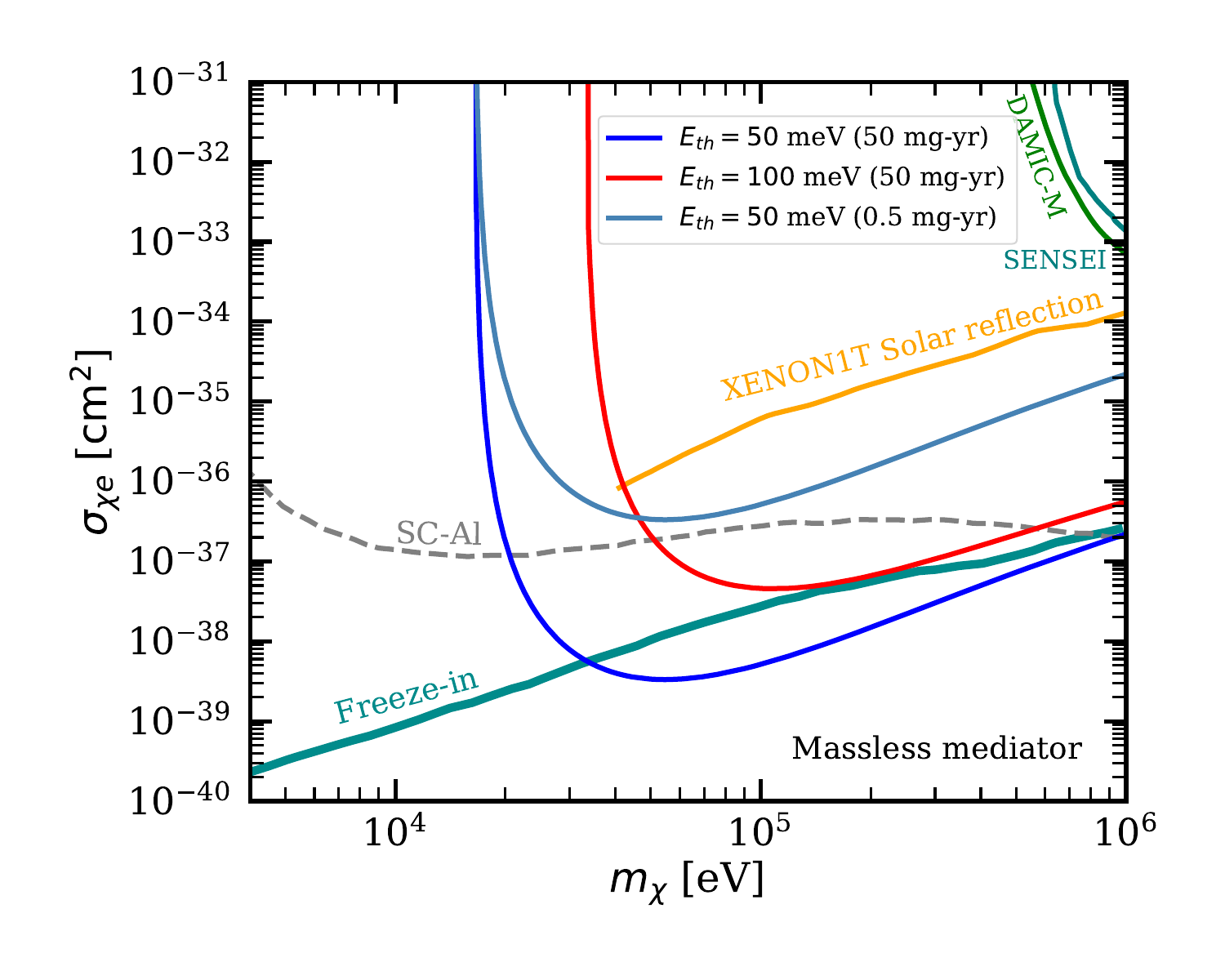}
    \includegraphics[width = 1.03 \columnwidth]{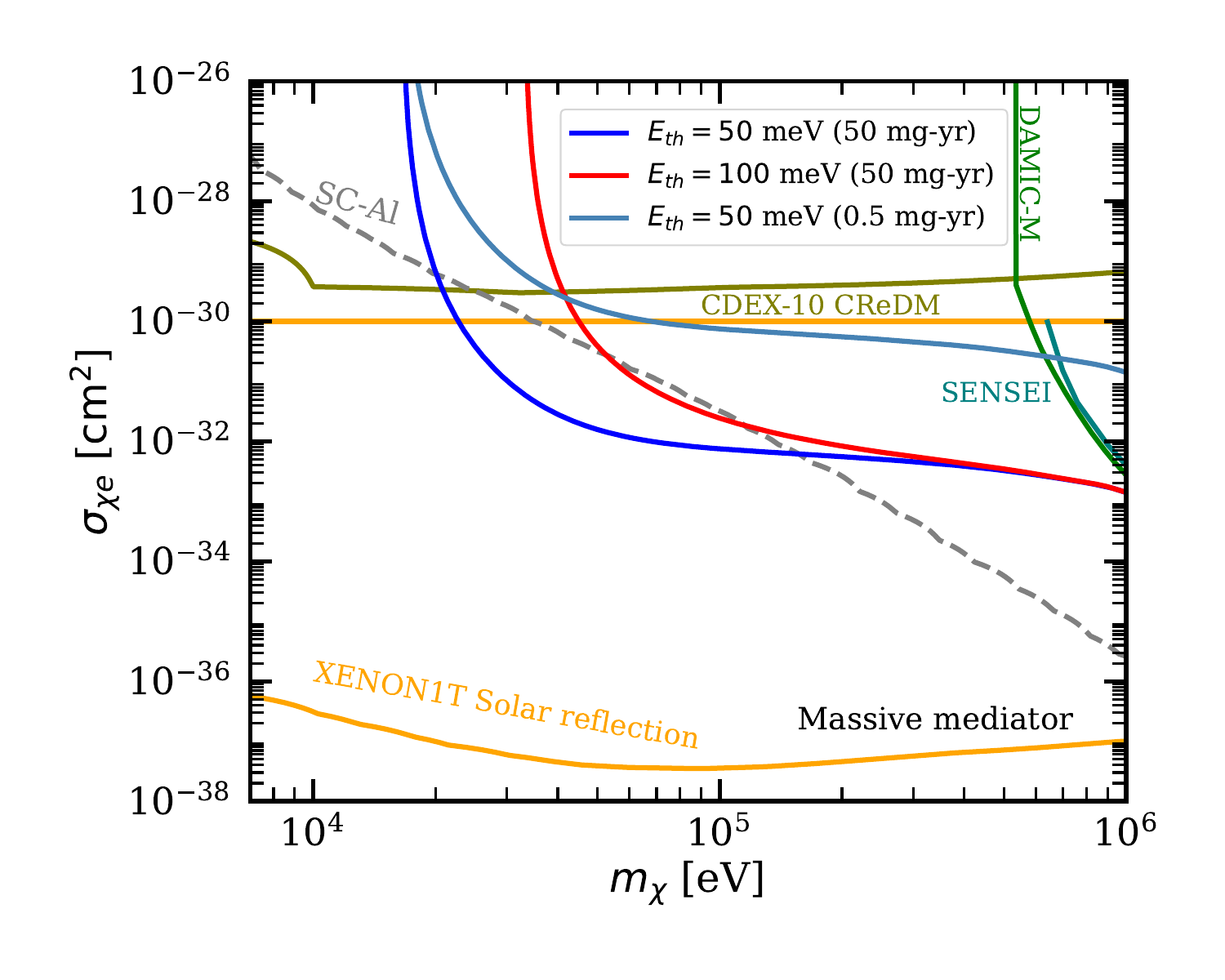}  
    \includegraphics[width =1.03\columnwidth]{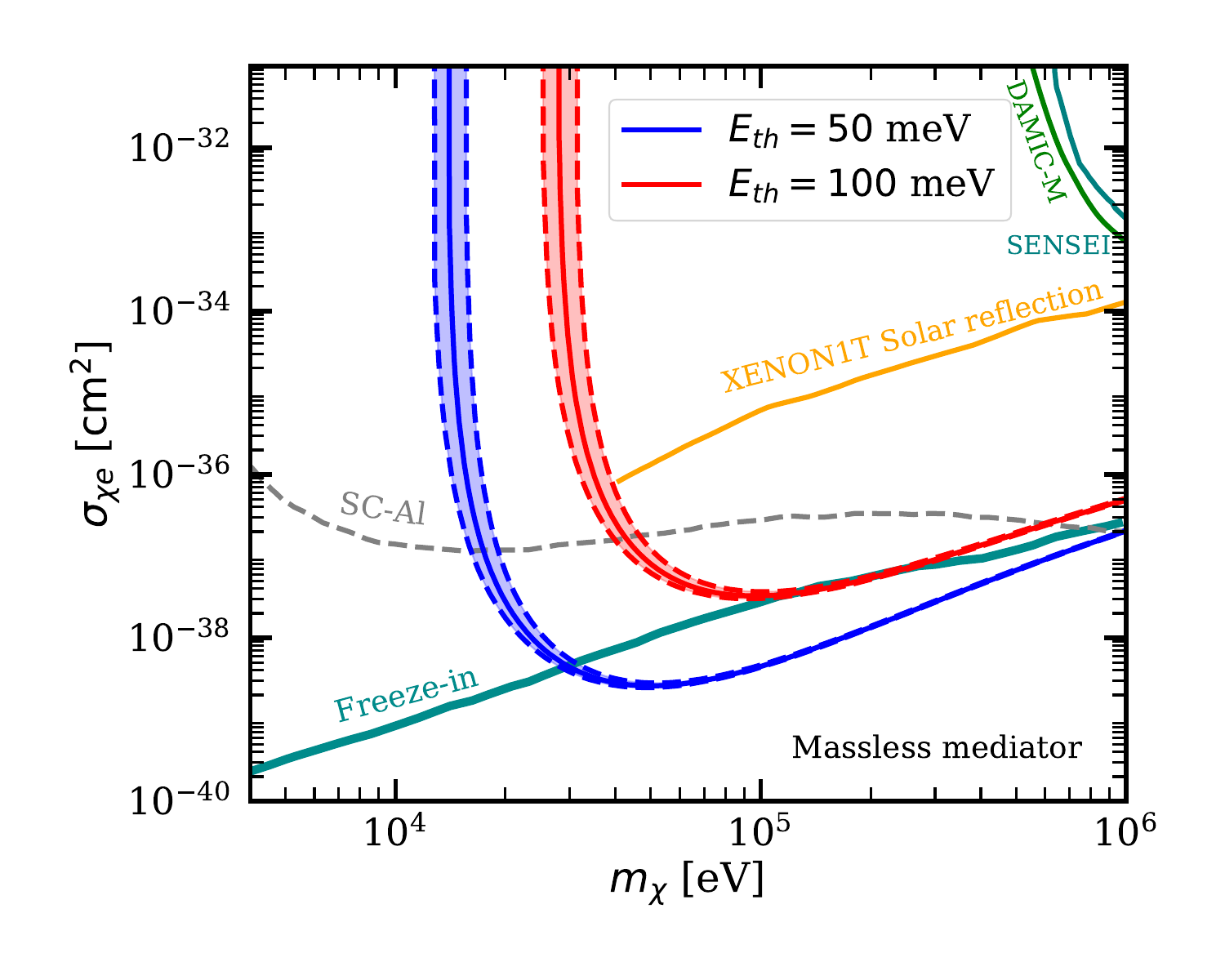     }
    \includegraphics[width =1.03 \columnwidth]{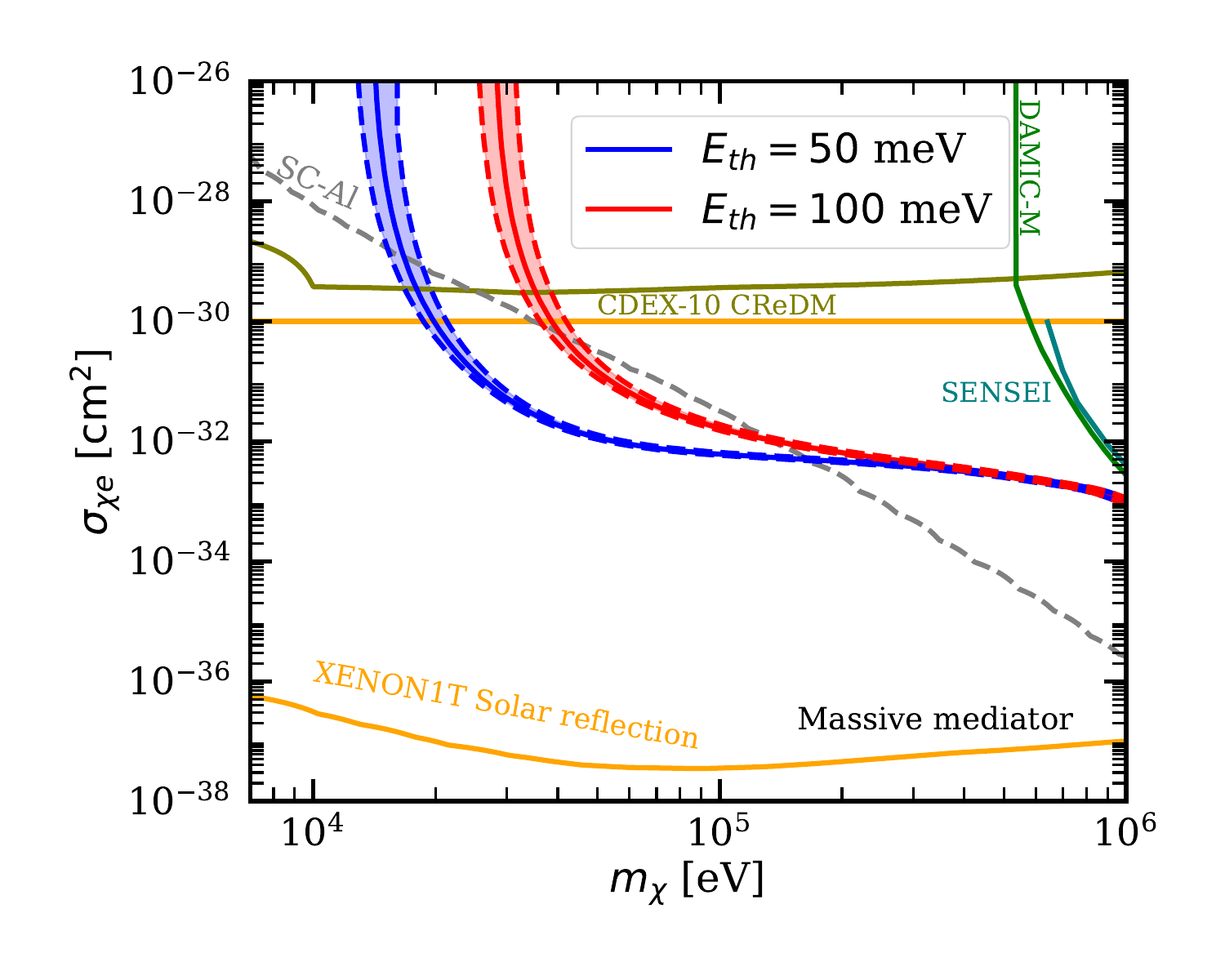}  
    \caption{Top:
    Projected sensitivity on the upper limit of DM-electron scattering $\sigma_{\chi e}$  via a massless mediator (left) and massive mediator (right) for $E_{\rm th}$ = 100 meV (red) and $E_{\rm th}$ = 50 meV (blue), assuming 3 events in 50 mg-year exposure and 0.5 mg-year exposure (steel blue) with zero background. For comparison, the projection for the superconducting aluminium (SC-Al) for 50 mg year exposure is shown as a grey, dotted line\,\cite{Hochberg:2021yud}. The limits from XENON1T on Solar reflected DM is shown in orange and the limits from CDEX-10 on cosmic ray electron boosted light DM (CReDM) is shown in olive\,\cite{Emken:2021lgc,Xu:2026dae}. The green and teal lines are the limits from DAMIC-M and SENSEI experiments,  respectively\,\cite{DAMIC-M:2023hgj, SENSEI:2025qvp}. The freeze-in DM benchmark is shown in dark cyan. Bottom: Same as top but with projected sensitivity using the velocity distribution parameters from Ref.\,\cite{Folsom:2025lly}. The solid line is obtained using the mean value of the parameters, while the dashed lines correspond to the maximum and minimum uncertainties. } 
    \label{fig: Projected sensitivity}
\end{figure*}

\begin{figure*}[!htbp]
    \centering
        \centering
        \includegraphics[width = 1.5\columnwidth]{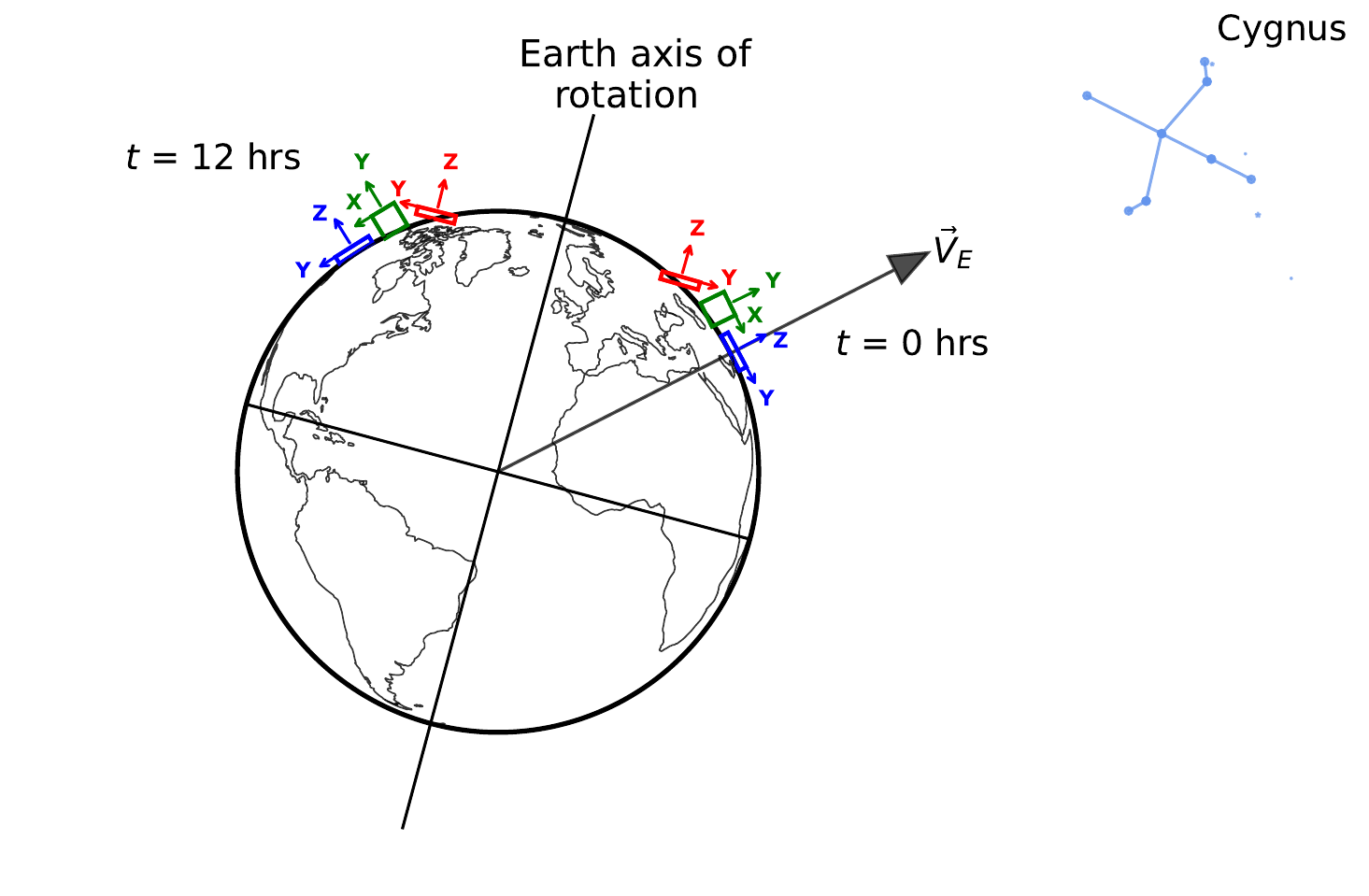} 
    \includegraphics[width=0.9\columnwidth]{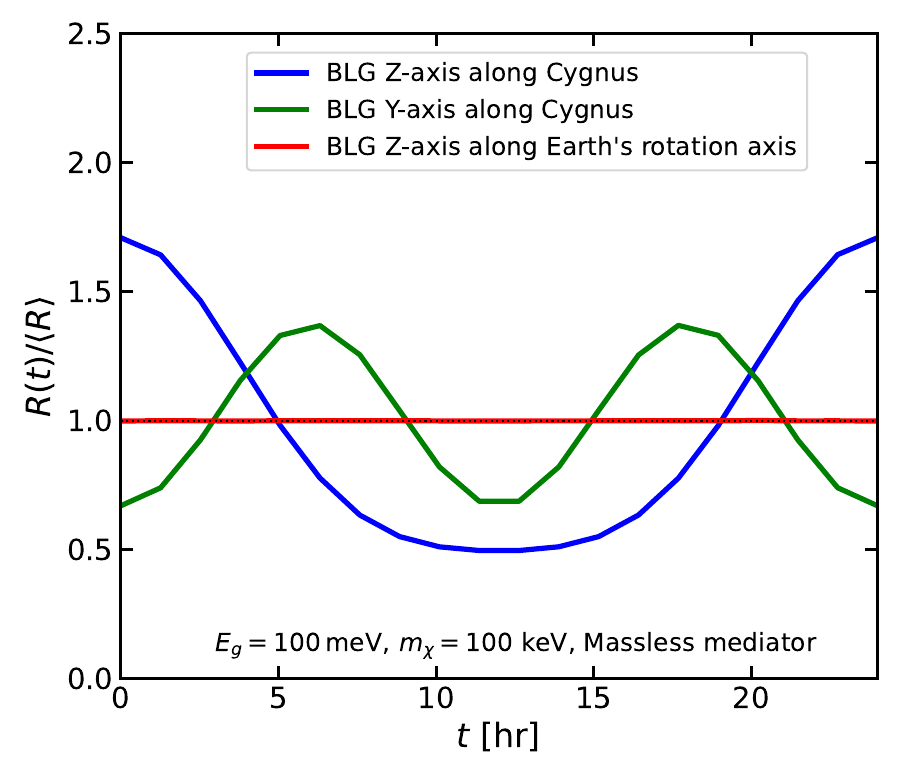}
    \includegraphics[width=0.9\columnwidth]{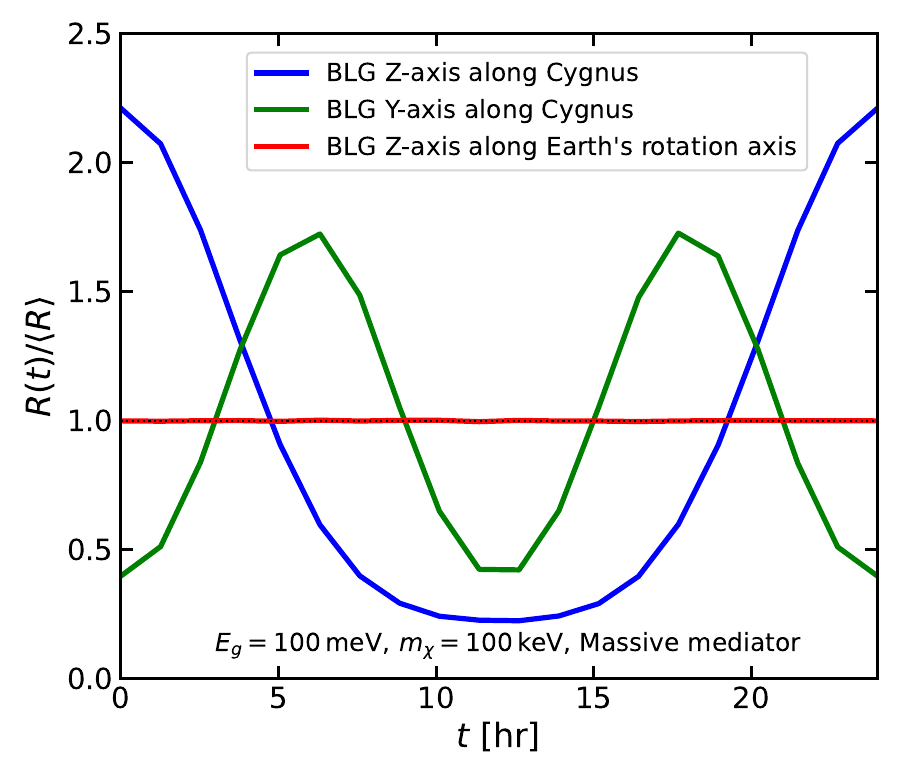}
    \caption{Illustration of rotation of the Earth and change of orientation of BLG with respect to DM wind (top). Effect of BLG orientation on daily modulation signal for massless (bottom left) and massive (bottom right) mediator. We considered three different orientations, with the Z-axis (blue, orientation-1), the Y-axis (green, orientation-2) of the BLG oriented along the direction of the galactic DM wind at $t$ = 0 hours, and when the Z-axis is oriented along the Earth's axis of rotation (red, orientation-3).}   
    \label{fig: daily modulation variation with Orientation}
\end{figure*}

\begin{figure*}[!htbp]
\centering
\includegraphics[width = 0.8\columnwidth]{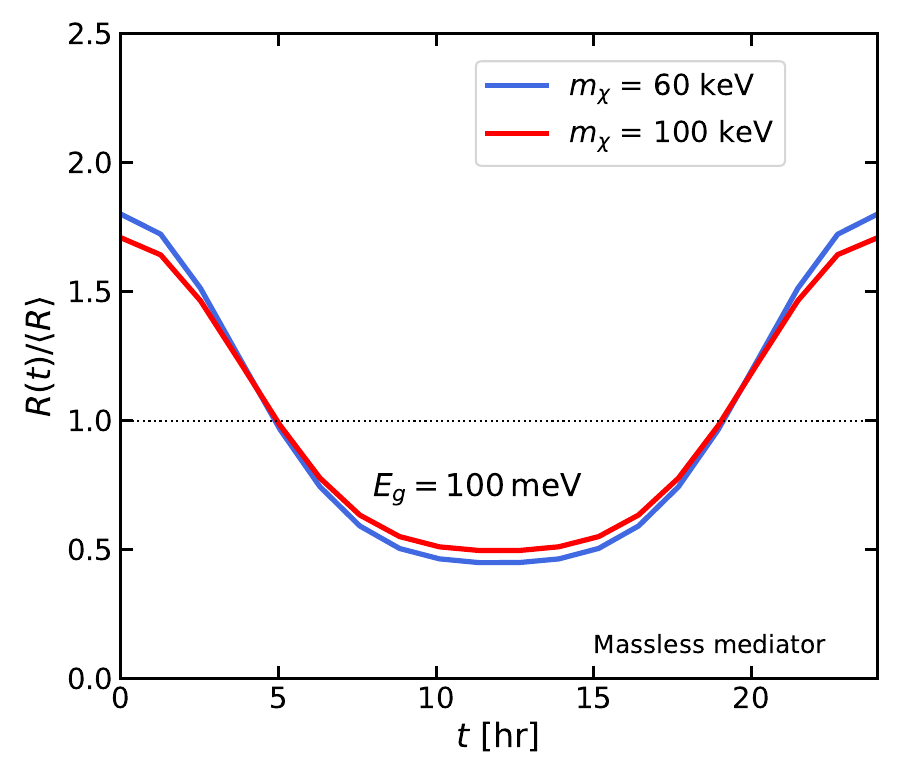}
\includegraphics[width = 0.8\columnwidth]{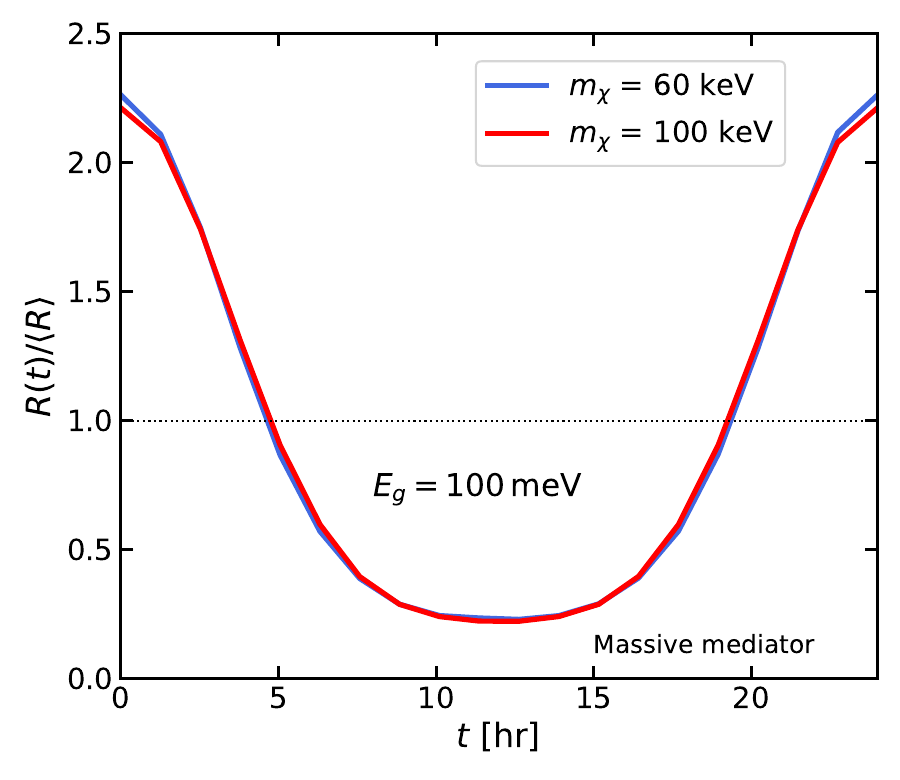}   
\caption{ {Daily modulation of DM-electron scattering rate via massless mediator (left) and massive mediator (right) in BLG for band gap $U$ = 100 meV and DM mass $m_\chi$ = 60 keV (blue) and $m_\chi$ = 100 keV (red) with the Z-axis of the BLG along the direction of Cygnus at $t$ = 0 hrs.}} 
\label{fig: daily modulation variation with mass}
\end{figure*}

\section{Energy loss function and Scattering rate} 
\label{sec: projected sensitivity}
Using eigenvalues and eigenvectors of the tight-binding Hamiltonian of Eq.~\,\eqref{equ: tight binding hamiltonian}, we calculate the integral in Eq.~\,\eqref{equ: non-interacting polarization function integral} over the first BZ to get the non-interacting polarization function, which in turn gives the ELF of BLG. The value of ELF characterizes the electronic energy loss at a given $\mathbf{q}$ and $\omega$. In BLG, this energy loss is due to electron excitations from the valence to conduction bands, and the value of ELF tells whether an excitation with $\mathbf{q}$ and $\omega$ is likely or not. As the Fermi energy lies within the band gap, there is no contribution to the ELF from intraband transitions. 

In Fig.~\ref{energy loss function}, we show the ELF as a function of in-plane momentum transfer $\mathbf{q}_{\parallel}$ for $\omega$ = 0.5 eV (top left), $\omega$ = 1 eV (top right), $\omega$ = 2 eV (middle left) and $\omega$ = 3 eV (middle right). In the bottom panel of  Fig.~\ref{energy loss function}, we show the ELF as a function of energy ($\omega$) and momentum transfer in the X-direction ($q_x$). The solid and dashed red lines indicate the kinematically allowed boundaries due to energy-momentum conservation, coming from the argument of the delta function in Eq.\,\eqref{equ: G - function definition}, for $m_\chi$ = 1 MeV and $m_\chi$ = 600 keV, respectively. In these figures, the region where ELF is non-zero is known as the electron-hole continuum. The ELF is highly peaked and isotropic at low values of $\mathbf{q}_{\parallel}$. This ELF peak mostly represents the low-energy and low-momentum
transfer transition states possible around the K points. We show such small momentum transitions with the letter `S' in Fig.~\ref{band structure} (left) and the corresponding region in the ELF where such transitions contribute the most in Fig.~\ref{energy loss function}. For larger momentum transfers, the energy difference between the valence bands and conduction bands increases,  and therefore, the value of ELF decreases if the transferred energy is small. The ELF again peaks anisotropically around the six corners of the hexagonal BZ, at very large values of $\mathbf{q}_{\parallel}$. These peaks correspond to transitions possible from the region around $K_+$ ($K_-$) point in the valence band to the region around the $K_-$ ($K_+$) point in the conduction band by a momentum transfer of about $q \sim 3.3$ keV along the six sides of the hexagon.
This anisotropy can be seen at lower momentum transfers if the transferred energy is also large enough. We show such large $\mathbf{q}$ transitions with the letter `L' in the right panel of Fig.~\ref{band structure} and the ELF region where they contribute the most in Fig.~\ref{energy loss function}. We use this ELF and set $F_{\text{DM}}(q) = q_0^2/q^2$ (assuming $m_{\phi} \ll q$) and $F_{\text{DM}}(q)=1$ (assuming $m_{\phi}\gg q$) in Eq.~\eqref{equ: differential scattering rate} to obtain the differential DM-electron scattering rate via massless and massive mediators, respectively. 

In Fig.~\ref{fig: differential scattering rate}, we show this differential scattering rate as a function of energy $\omega$, for massless (left) and massive mediator (right), for time $t= 1$\, hr, for  $U$ = 100 meV, $\sigma_{\chi e}$ = $10^{-38}\,\mathrm{cm^2}$, $m_\chi$ = 100 keV (blue), 200 keV (orange) and 300 keV (green). The non-zero rate below the energy threshold, in Fig.~\ref{fig: differential scattering rate}, is due to the finite value of decay width $\eta$ of the excited electron-hole pairs. Note that the rate for the massive mediator is much lower than that for the massless mediator, keeping everything else fixed.

The projected sensitivity for massless (left) and massive (right) mediators, assuming 3 events in 50 mg-year exposure with zero background, is shown for $E_{th}$ = 50 meV (blue) and $E_{th}$ = 100 meV (red) in Fig.~\ref{fig: Projected sensitivity} top panel. Also, the projected sensitivity for a much lower exposure of 0.5 mg-year is also shown for $E_{th}$ = 50 meV in steel blue in Fig.~\ref{fig: Projected sensitivity}. The projected sensitivity using the velocity distribution parameters suggested by Ref.\,\cite{Folsom:2025lly} is shown in the bottom panel of Fig.~\ref{fig: Projected sensitivity}, with the solid line denoting the mean parameter values. The dashed lines correspond to the maximum and minimum uncertainties in these parameters coming from the simulation\,\cite{Folsom:2025lly}. For comparison, the projection for the superconducting aluminium (SC-Al) for 50 mg-year exposure is shown as a
gray, dotted line\,\cite{Hochberg:2021yud}. The limits from XENON1T on solar reflected DM is shown in orange and the limits from CDEX-10 on cosmic ray electron boosted light DM (CReDM) is shown in olive~\cite{Emken:2021lgc,Xu:2026dae}. The green and teal lines are the limits from DAMIC-M and SENSEI experiments,  respectively\,\cite{DAMIC-M:2023hgj, SENSEI:2025qvp}. The dark cyan line indicates the cross-section and mass range where DM can be produced via the freeze-in mechanism, which BLG can probe with 50 mg-year exposure. Note that even if we lower the exposure to $\sim$ 0.5 mg-year, BLG will still be able to probe new regions of the parameter space. Our sensitivity to DM–electron scattering with $F_{\rm DM}=1$ does not follow the usual trend where the reach degrades with increasing DM mass, instead, it slightly improves as the DM mass increases (in the DM mass range that we consider). This arises from the combined effect of uniform weightage from $F_{\rm DM}$ over all momentum transfers, together with the increase in maximum energy transfer as the DM mass grows, which in turn opens up a larger accessible phase space. In contrast, in the massless mediator case, the smallest momentum transfers receive the largest weightage, and such low-momentum transfers can be produced by all DM particles above a certain mass set by the detector threshold.

\section{Daily Modulation of Scattering rate} 
\label{sec:Dailymod}
The Sun, along with the Earth, is moving towards the direction of the Cygnus constellation in the Galactic rest frame, see Fig.~\ref{fig: daily modulation variation with Orientation}. This motion creates a DM wind with a unique direction in the laboratory frame, which, if detected, will be a telltale signature for  DM. Directional detectors try to do this by looking at the recoil direction of the nucleus after scattering with the DM particle. Whilst this can be done to detect the direction of heavy DM particles, it is not feasible to do so for less energetic light DM.
In such case, the daily modulation of the scattering rate in time due to anisotropy in the detector as the direction of DM wind rotates in the lab frame, will be an important handle to identify DM-induced events and reject background (for other sources of daily modulation see e.g., Ref.\,\cite{SENSEI:2025qvp,DAMIC-M:2025ltz}). Most of the DM particles impart momentum along the direction of the DM wind. The structure factor or equivalently ELF tells whether a transition at a particular momentum and energy transfer is possible or not. The 3D structure factor of BLG is a product of in-plane 2D structure factor and a $q_z$-dependent part, which is simply the square of a sinc function. For the masses we have considered, the value of the $q_z$-dependent sinc function is almost constant for all momentum transfers. 

When the DM wind is perpendicular to the BLG plane, most of the DM particles transfer their momentum along the Z-direction, and very little momentum is imparted in the X or Y direction. Since ELF as shown in Fig.~\ref{energy loss function}, is peaked at small values of $q_x$ and $q_y$ we get a large scattering rate.  When the angle between the DM wind and BLG plane is less than $90^\circ$, there is a larger momentum transfer in the X-Y plane, where ELF has a lower value, and thus we find a smaller rate.  This is the cause of the modulation signal in BLG. The anisotropy in ELF at large momentum $\mathbf{q}_{\parallel}$ and energy transfer $\omega$ did not contribute to any additional features in modulation, for DM masses and velocities that we have considered. This is because of the non-overlap of the kinematically allowed region for DM and the anisotropic part of the electron-hole continuum, as shown in Fig.~\ref{energy loss function} bottom panel. For heavier mass DM or velocity boosted DM, this anisotropy could contribute to the rate and modulation.

The rate modulation shape depends on the orientation of the BLG with respect to the DM wind. Different orientations correspond to different $\vec{v}_{\text{lab}}(t)$, in Eq.\,\eqref{equ: G - function analytic form}. Neglecting the small variation of Earth's speed due to rotation around its axis and revolution around the Sun in a day, the lab velocity can be written as,
\begin{equation}
\label{equ: lab velocity  Vlab}
    \vec{v}_{\text{lab}}(t)= |\mathbf{v}_E|\hat{\bold{v}}_e(t)
\end{equation}
where $|\mathbf{v}_E|$ = 240 km/s and $\hat{\bold{v}}_e(t)$ is the time-dependent unit vector that is pointing towards the direction of  Cygnus constellation in the BLG coordinate system \cite{Griffin:2018bjn}.

In order to see the effect of BLG orientation on the modulation signal, we consider three different orientations of BLG with respect to the DM wind, as illustrated in Fig.~\ref{fig: daily modulation variation with Orientation}. In the first orientation (orientation-1), the Z-axis of the BLG is along the DM wind direction at $t = 0$ hrs (shown in blue in Fig.~\ref{fig: daily modulation variation with Orientation}).  In the second orientation (orientation-2), the $\mathrm{X-Y}$ plane of the BLG is along the DM wind direction at $t = 0$ hrs (shown in green in Fig.~\ref{fig: daily modulation variation with Orientation}) and in the third orientation (orientation-3), the $\mathrm{Z}$-axis of the BLG is along the Earth's axis of rotation (shown in red in Fig.~\ref{fig: daily modulation variation with Orientation}).
Now, $\hat{\bold{v}}_e(t)$ for orientation-3, as a function of time can be written as
\begin{equation}
\label{equ: lab velocity orientation-1}
\hat{\bold{v}}_e(t) =  
    \begin{pmatrix}
        \sin\theta_e \sin\phi(t) \\
        \sin\theta_e \cos\phi(t) \\
        \cos\theta_e
    \end{pmatrix}.
\end{equation}
where, 
$\phi(t) = 2\pi \times \left( \frac{t}{24\,\text{h}} \right)\, \text{and} \quad 
\theta_e = 42^\circ$ (angle between the Earth's axis of rotation and the direction of Cygnus).
To obtain $\hat{\bold{v}}_e(t)$ for orientation-1, we rotate $\hat{\bold{v}}_e(t)$ of Eq.\,\eqref{equ: lab velocity orientation-1} by $\theta_e$ about the X-axis. The active rotation is performed using Rodrigues' rotation formula, which gives the rotated vector $\mathbf{v}_{\rm rot}$ obtained by rotating a vector $\vec{v}$ about an arbitrary unit vector $\hat{\bold{k}}$ by angle $\theta_{\rm rot}$ as,
\begin{equation}
\label{equ: Rodrigues' rotation formula}
\mathbf{v}_{\text{rot}} = 
\mathbf{v} \cos\theta_{\rm rot} 
+ (\hat{\bold{k}} \times \mathbf{v}) \sin\theta_{\rm rot} 
+ \hat{\bold{k}}(\hat{\bold{k}} \cdot \mathbf{v})(1 - \cos\theta_{\rm rot}).
\end{equation}
Substituting the above expressions for $\mathbf{v}$ = $\hat{\bold{v}}_e(t)$, $\hat{\bold{k}}$ = $\mathbf{X}$, and $\theta_{\rm rot}$ = $\theta_e$ and simplifying, we obtain the following equation,
\begin{equation}
\label{equ: lab velocity orienation -1}
\hat{\bold{v}}_e(t) =  
  \begin{pmatrix}
\sin\theta_e \sin\phi(t) \\
\sin\theta_e \cos\theta_e (\cos\phi(t) - 1) \\
\cos^2\theta_e + \sin^2\theta_e \cos\phi(t)
\end{pmatrix}.
\end{equation}
Similarly, to get $\hat{\bold{v}}_e(t)$ for orientation-2, we rotate $\hat{\bold{v}}_e(t)$ of Eq.\,\eqref{equ: lab velocity orientation-1} by $-(\pi/2-\theta_e)$ about the X-axis followed by a rotation about the Y-axis by $-\pi/2$ and get
\begin{equation}
\label{equ: lab velocity orienation -2}
\hat{\bold{v}}_e(t) =
\begin{pmatrix}
 \sin\theta_e \cos\theta_e(\cos\phi(t) -1) \\
\cos^2\theta_e +\sin^2\theta_e \cos\phi(t) \\
\sin\theta_e \sin\phi(t)
\end{pmatrix}.
\end{equation}

In orientation-1 (shown in blue in Fig.~\ref{fig: daily modulation variation with Orientation}), corresponding to $\hat{\bold{v}}_e(t)$ of Eq.\,\eqref{equ: lab velocity orienation -1} in the BLG frame, the BLG Z-axis is aligned with the DM wind at $t=0$\,hr and $t = 24$\,hr, and thus we get the maximum rate at these times. As the Earth rotates, the angle between the BLG plane and the DM wind becomes less than $90^\circ$, resulting in a lower rate. To see the effect of the DM mass, we also plot the daily modulation of scattering rate for orientation-1 with threshold energy $E_{\rm th}$ = 100 meV, and DM masses $m_\chi$ = 60 keV (blue) and $m_\chi$ = 100 keV (red) in Fig.\,\ref{fig: daily modulation variation with mass} (massless (left) and massive mediator (right)). For smaller DM masses, a larger region of the kinematically allowed phase space may fall below the threshold energy compared to heavier DM, resulting in a larger modulation for smaller DM masses. When the DM wind is perpendicular to the BLG, although $q_x$ and $q_y$ values are small, the $  q_z$ component and therefore the magnitude $q$ can still be large, and the rate can be suppressed by $1/q^4$ when the mediator is massless. This, therefore, suppresses the relative modulation in the massless mediator's scattering rate compared to the massive mediator's scattering rate, as seen in Fig.~\ref{fig: daily modulation variation with mass}. 

In orientation-2 (shown in green in Fig.~\ref{fig: daily modulation variation with Orientation}), corresponding to the $\hat{\bold{v}}_e(t)$ Eq.\,\eqref{equ: lab velocity orienation -2}, the $\mathrm{X-Y}$ plane of the BLG is oriented along the DM wind direction at $t = 0$ hrs. In this orientation, we start with a lower rate as there is a larger momentum transfer in the $\mathrm{X-Y}$ plane, where ELF has a lower value. The rate then increases, reaching a maximum at around 6 hrs when the BLG plane is facing the DM wind, and then decreases, and again reaches a maximum at about 18 hrs when the BLG plane is again facing the DM wind. 

In orientation-3 (shown in red in Fig.~\ref{fig: daily modulation variation with Orientation}), the Z-axis of the BLG is along the direction of the Earth's rotation axis, and corresponds to $\hat{\bold{v}}_e(t)$ (see Eq.\,\eqref{equ: lab velocity orientation-1}). As can be seen from Fig.~\ref{fig: daily modulation variation with Orientation}, there is no modulation in this configuration. This is because the DM wind flux incident on the BLG plane does not change with Earth's rotation in this orientation. 

As discussed, the modulation signature is regulated by $g(\vec{q}, \omega, t)$ and the anisotropic structure factor $S(\vec{q}, \omega)$ for a given orientation of BLG. To illustrate this further in appendix \ref{app:daily_modulation} we show the time variation of $g(\vec{q}, \omega, t)$ in Figs.\,\ref{fig: appendix Gfunction 1} and \ref{fig: appendix Gfunction 2}  while the lower panel of Fig.\,\ref{fig: appendix Gfunction 2} shows time independent $S(\vec{q}, \omega)$. When the peak of $g(\vec{q}, \omega, t)$ coincides with the peak of  $S(\vec{q}, \omega)$ the event rate is maximized, conversely, when they do not overlap, the rate is suppressed.

We note the importance of the orientation of the anisotropic detectors like BLG, since some orientations, like the orientation-1, are ideal for dark matter detection, as they provide the maximum modulation, whereas other orientations might give little or no modulation at all, like the orientation-3 in our case. The daily modulation signal is of paramount importance to identify and confirm any telltale signature of DM in an experiment, especially in the presence of background events of unknown source\,\cite{Baxter:2025odk}. Even though all sources of the background events are not known, they are not expected to share the same amplitude and phase of the daily modulation as the DM signal. Hence, the modulation signal, as explained here, could distinguish between DM and background events. Anisotropic targets like BLG are appropriate for this purpose.

\section{Conclusions}
\label{sec:conclusions}
In this work, we have calculated the scattering rate of DM particles with electrons in BLG via a massless and massive mediator. For this, we have used a tight-binding model of BLG to calculate the non-interacting polarization function, using which we have then calculated the ELF of BLG in RPA. We have projected the sensitivity of BLG for a 50 mg-year exposure without background, and we notice that BLG provides competitive limits, similar to other materials, such as superconducting aluminium\,\cite{Hochberg_2021}. For a massless mediator, BLG is able to probe the freeze-in benchmark with this level of exposure. Note that BLG will still be able to probe new regions of the parameter space with a much lower exposure of $\sim$ 0.5 mg-year. We have then looked at the daily modulation of scattering rate due to anisotropies in the BLG structure factor and the dependence of this modulation signal shape on BLG configuration. This modulation is coming from the $\mathrm{X-Y}$ and $\mathrm{Z}$ asymmetry in the structure factor. The $\mathrm{X-Y}$ structure factor/ELF has asymmetry at higher momentum, but this does not affect the DM masses that we have considered due to the non-overlap of kinematically allowed DM phase space and electron-hole continuum.

The tight-binding model used in our work is strictly not valid over the entire BZ as parameters are fitted only near the K points\,\cite{McCann_2011}. The actual band structure around the $\Gamma$ point may differ from that used in our calculations. Furthermore, there may be additional contributions from other bands, which can be particularly important for heavier DM masses\,\cite{PhysRevB.93.235153DFTBLG}. We believe these factors would not have much effect on our results as the kinematically allowed region for DM masses that we have considered only overlaps at low $q$ and $\omega$ region of the electron-hole continuum, and even if the true electron-hole continuum overlaps significantly at large $q$ and $\omega$, that would only increases the rate further; in that case, our results can be considered as a conservative estimate. Our calculations considered only $\pi$ electrons and restricted momentum transfers below 3 keV (within the first BZ). For higher mass DM or velocity boosted DM, the $\mathrm{X-Y}$ anisotropy in the structure factor could contribute to the rate. If the energy of the DM is greater than the BLG's work function, the electron could be ejected out, similar to Ref.\,\cite{Wang:2015kya, Hochberg:2016ntt, Catena:2023awl}. The design of the BLG detector has been discussed in Ref.\,\cite{Das:2023cbv}.

As with all the low-threshold experiments, scaling the detector and achieving the required exposure is a challenge with BLG. Current chemical vapour deposition techniques report single AB stacked BLG crystal of lateral size $\mathcal{O}(100)\,\mu\mathrm{m}$\,\cite{Zhou2013}.
However, there have been works where people have produced much larger single layers of graphene\,\cite{puregraph,Nguyen2020,patent_Jeong}. In future, we might be able to produce much larger size graphene layer/bilayer and it might be possible to reach the $\mathcal{O}(10)\,\mathrm{mg}$ exposure by building multiple modules as discussed in Ref.\,\cite{Das:2023cbv}. This may lead us to the discovery of the nature of dark matter.

\textit{Acknowledgments--}
The authors thank Adarsh Abraham Basumata,  Debajit Bose, Subhadip Bouri, Deep Jyoti Das, Hiranmay Das, Basudeb Dasgupta, Phanindra Dewan, Pugazhendhi A.\,D., Jaya Doliya, Abhishek Dubey, Durba Ghosh, Jiho Jang,  Udit Khanna, Ranjini Mondol, Akash Kumar Saha, Diptiman Sen, Rajdeep Sensarma and Abhijeet Singh for helpful discussions.
R.S. acknowledges the University Grants Commission (UGC), Government of India, for the financial support via the UGC-NET Senior Research Fellowship. The work of T.N.M is supported by the Australian Research Council through the ARC Centre of Excellence for Dark Matter Particle Physics. The work by P.D. at Physical Research Laboratory is supported by the Department of Space (DoS), Government of India. P.D. acknowledges the International Centre for Theoretical Sciences (ICTS) program - Engineered 2D Quantum Materials (code: ICTS/E2QM2024/07) for stimulating discussions on bilayer graphene.  R.L. and P.D. acknowledge the financial support from the IISc-ISRO STC (grant no.~ISTC/PHY/RL/499). R.L. also acknowledges ANRF for the financial support through ARG (grant no. ANRF/ARG/2025/005140/PS). A.D. acknowledges ANRF for the financial support through PMECRG (Grant no. ANRF/ECRG/2025/001012/PMS).

\appendix
\section{Tight Binding Model}
\label{app: tight binding model}
In this section, we show the derivation of the tight-binding Hamiltonian. Following Ref.\,\cite{McCann_2011}, the Bloch functions with wave-vector $\textbf{k}$ at position $\textbf{r}$ can be written as,
\begin{equation}
\label{equ:3}
    \Phi_j(\mathbf{k}, \mathbf{r}) = \frac{1}{\sqrt{N}} \sum_{i=1}^{N} e^{i \mathbf{k} \cdot \mathbf{R}_{j,i}} \, \phi_j\left( \mathbf{r} - \mathbf{R}_{j,i} \right),
\end{equation}
where $\phi_j\left( \mathbf{r} - \mathbf{R}_{j,i} \right)$  are the $\pi$ atomic orbitals, $\mathbf{R}_{j,i}$ is the position vector of $j^{th}$ orbital (the four orbitals labeled by $A_1,A_2,B_1,B_2$ for BLG) in $i^{th}$ unit cell and $N$ is the number of unit cells.
The general electronic wavefunctions can be written as a linear combination of these Bloch functions,
\begin{equation}
\label{equ:4}
   \Psi_n(\mathbf{k}, \mathbf{r}) = \sum_{j} c_{nj}(\mathbf{k}) \, \Phi_j(\mathbf{k}, \mathbf{r}) 
\end{equation}
with the energy of the $n^{th}$ band given by, 
\begin{equation}
\label{equ:5}
E_n(\mathbf{k}) = \frac{\langle \Psi_n | \mathcal{H} | \Psi_n \rangle}{\langle \Psi_n | \Psi_n \rangle},
\end{equation}
 where $\mathcal{H}$ is the Hamiltonian. Minimizing the energy with respect to the coefficient $c_{nm}(\textbf{k})$ gives the following equation,
\begin{equation}
\label{equ:6}
  H \psi_n = E_n \mathcal{S} \psi_n .
\end{equation}
where $\psi_n$ is a column vector with $(\psi_n)_l=c_{nl}(\textbf{k})$; $H$ and $S$ are matrices with elements  
\begin{equation}
\label{equ:Hmatrix}
\begin{aligned}
H_{il} &= \langle \Phi_i | \mathcal{H} | \Phi_l \rangle \\
&= \frac{1}{N} \sum_{m=1}^{N} \sum_{j=1}^{N} e^{i \mathbf{k} \cdot (\mathbf{R}_{l,j} - \mathbf{R}_{i,m})} \\
&\quad \times \left\langle \phi_i (\mathbf{r} - \mathbf{R}_{i,m}) \middle| \mathcal{H} \middle| \phi_l (\mathbf{r} - \mathbf{R}_{l,j}) \right\rangle,
\end{aligned}
\end{equation}

\begin{equation}
\label{equ:Smatrix}
\begin{aligned}
\mathcal{S}_{il} &= \langle \Phi_i | \Phi_l \rangle \\
&= \frac{1}{N} \sum_{m=1}^{N} \sum_{j=1}^{N} e^{i \mathbf{k} \cdot (\mathbf{R}_{l,j} - \mathbf{R}_{i,m})} \\
&\quad \times \left\langle \phi_i (\mathbf{r} - \mathbf{R}_{i,m}) \middle| \phi_l (\mathbf{r} - \mathbf{R}_{l,j}) \right\rangle.
\end{aligned}
\end{equation}
We consider only the onsite contribution for diagonal terms and the nearest-neighbor hopping contribution for off-diagonal terms in Eqs.\,\eqref{equ:Hmatrix} and\,\eqref{equ:Smatrix}, as other contributions have a very small effect on the band structure.  Moreover, we do not consider off-diagonal terms in $\mathcal{S}$ as including these would modify the band structure near the $\Gamma$ point, but that would have a negligible contribution to the scattering rate since the dominant contributions come from the region at and around the $K$ points. Therefore, $\mathcal{S}$ becomes an identity matrix and the diagonal terms of $H$ are,
\begin{align}
    H_{jj} &\approx \frac{1}{N} \sum_{i=1}^{N}
\left\langle \phi_j(\mathbf{r} - \mathbf{R}_{j,i})
\middle| \mathcal{H} \middle|
\phi_j(\mathbf{r} - \mathbf{R}_{j,i})
\right\rangle\nonumber\\
&\approx \left\langle \phi_j
\middle| \mathcal{H} \middle|
\phi_j
\right\rangle = \varepsilon_j
\end{align}
and the off-diagonal terms of $H$ are,
\begin{widetext}
\begin{eqnarray}
 H_{A_1 A_2} &\approx & \frac{1}{N}
\sum_{i=1}^{N} \sum_{l=1}^{3}
e^{i \mathbf{k} \cdot (\mathbf{R}_{A_2,l} - \mathbf{R}_{A_1,i})}
\left\langle
\phi_{A_1}(\mathbf{r} - \mathbf{R}_{A_1,i})
\middle| \mathcal{H} \middle|
\phi_{A_2}(\mathbf{r} - \mathbf{R}_{A_2,l})
\right\rangle \nonumber\\
&\approx & \frac{1}{N}
\sum_{i=1}^{N} \sum_{l=1}^{3}
e^{i \mathbf{k} \cdot (\mathbf{R}_{A_1,i}+\mathbf{\Delta}_l - \mathbf{R}_{A_1,i})} \left\langle
\phi_{A_1}(\mathbf{r} - \mathbf{R}_{A_1,i})
\middle| \mathcal{H} \middle|
\phi_{A_2}(\mathbf{r} - \mathbf{R}_{A_2,l})
\right\rangle \nonumber\\
&\approx & \frac{1}{N}
\sum_{i=1}^{N} \sum_{l=1}^{3}
e^{i \mathbf{k} \cdot \bold{\Delta}_l} \left\langle
\phi_{A_1}(\mathbf{r} - \mathbf{R}_{A_1,i})
\middle| \mathcal{H} \middle|
\phi_{A_2}(\mathbf{r} - \mathbf{R}_{A_2,l})
\right\rangle\nonumber \\
&\approx & \left\langle
\phi_{A_1}
\middle| \mathcal{H} \middle|
\phi_{A_2} \right\rangle
\sum_{l=1}^{3}
e^{i \mathbf{k} \cdot \bold{\Delta}_l}\nonumber\\
& \approx & \gamma_4 f(\mathbf{k})
\end{eqnarray}
\end{widetext}
where $\mathbf{\Delta}_1 = \left( 0, \dfrac{a}{\sqrt{3}} \right)$, $
\mathbf{\Delta}_2 = \left( \dfrac{a}{2},-\dfrac{a}{2\sqrt{3}} \right)$, 
$\mathbf{\Delta}_3 = \left( -\dfrac{a}{2}, -\dfrac{a}{2\sqrt{3}} \right)$ and 
\begin{align}
f(\mathbf{k})
&= \sum_{l=1}^{3} e^{i \mathbf{k} \cdot \boldsymbol{\Delta}_l} \nonumber\\
&= e^{i k_y a/\sqrt{3}}
+ e^{i k_x a/2} e^{-i k_y a/2\sqrt{3}}
+ e^{-i k_x a/2} e^{-i k_y a/2\sqrt{3}} \nonumber\\
&= e^{i k_y a/\sqrt{3}}
+ 2 e^{-i k_y a/2\sqrt{3}} \cos\!\left(\frac{k_x a}{2}\right).
\end{align}
Similarly, we have
\begin{align}
H_{B_1,B_2}&\approx\left\langle
\phi_{B_1}
\middle| \mathcal{H} \middle|
\phi_{B_2}
\right\rangle f(\mathbf{k})=\gamma_4 f(\mathbf{k})\\
H_{A_1 B_1}&\approx \langle \phi_{A_1} | \mathcal{H} | \phi_{B_1} \rangle f(\mathbf{k})
          = -\gamma_0f(\mathbf{k}) \\
H_{A_2 B_2}&\approx \langle \phi_{A_2} | \mathcal{H} | \phi_{B_2} \rangle f(\mathbf{k})
          = -\gamma_0f(\mathbf{k})\\
H_{A_2,B_1}&\approx \langle \phi_{A_2} | \mathcal{H} | \phi_{B_1} \rangle=\gamma_1 \\
H_{A_1,B_2}&\approx \langle \phi_{A_1} | \mathcal{H} | \phi_{B_2} \rangle f^*(\mathbf{k})=-\gamma_3 f^*(\mathbf{k}) 
\end{align}
\section{Derivation of Scattering rate}
\label{app: derivation of scattering rate}
We consider DM particles scattering off electrons in a target material via interactions mediated by a massive vector or scalar boson with mass $m_\phi$. The coupling constants are $g_\chi$ (DM-mediator) and $g_e$ (electron-mediator).
The non-relativistic DM-electron interaction Hamiltonian in the momentum space is:
\begin{equation}
\Delta H_{\chi e} = \sum_{j} \int \frac{d^3q}{(2\pi)^3} \frac{g_e g_\chi}{q^2 + m_\phi^2} e^{i\vec{q}\cdot(\vec{r}_{e,j} - \vec{r}_\chi)}
\end{equation}
where $\vec{r}_{e,j}$ is the position vector of the $j$-th electron and summation is over all the electrons in the target and $\vec{r}_\chi$ is the DM particle position operator. We calculate the transition amplitude from the 
initial state $|i,\vec{p}_{\chi}\rangle$, which is product of intial target state $|i\rangle$ and initial DM momentum state $|\vec{p}_\chi\rangle= e^{-i\vec{p}_\chi \cdot \vec{r}_\chi}/\sqrt{V}$ with momentum $\vec{p}_\chi$ to the final state  $|f,\vec{p}_{\chi}\rangle$,  which is the product of final target state $|f\rangle$ and the final DM momentum state $|\vec{p}_\chi^{'}\rangle= e^{-i\vec{p}'_\chi \cdot \vec{r}_\chi}/\sqrt{V}$ with momentum $\vec{p}'_\chi$. The matrix element is:
\begin{widetext}
\begin{eqnarray}
\langle f, \vec{p'}_{\chi} | \Delta H_{\chi e} | i, \vec{p}_{\chi} \rangle &=& \int \frac{d^3r_\chi}{V} \langle f | e^{-i\vec{p}'_\chi \cdot \vec{r}_\chi} \Delta H_{\chi e} e^{i\vec{p}_\chi \cdot \vec{r}_\chi} | i \rangle \nonumber \\
&=& \int \frac{d^3r_\chi}{V} \sum_{j} \int \frac{d^3q}{(2\pi)^3} \frac{g_e g_\chi}{q^2 + m_\phi^2} \langle f | e^{-i\vec{p}'_\chi \cdot \vec{r}_\chi} e^{i\vec{q}\cdot(\vec{r}_{e,j} - \vec{r}_\chi)} e^{i\vec{p}_\chi \cdot \vec{r}_\chi} | i \rangle \nonumber \\
&=& \sum_{j} \int \frac{d^3q}{(2\pi)^3} \frac{g_e g_\chi}{q^2 + m_\phi^2} \left[ \int \frac{d^3r_\chi}{V} e^{-i(\vec{q}-(\vec{p}_\chi - \vec{p}'_\chi))\cdot\vec{r}_\chi} \right] \langle f | e^{i\vec{q}\cdot\vec{r}_{e,j}} | i \rangle. \nonumber \\
&=& \int \frac{d^3q}{V} \delta^3(\vec{q} - (\vec{p}_\chi - \vec{p}'_\chi)) \frac{g_e g_\chi}{q^2 + m_\phi^2} \langle f | \sum_je^{i\vec{q}\cdot\vec{r}_{e,j}} | i \rangle  \nonumber \\
&=& \frac{1}{V} \frac{g_e g_\chi}{(\vec{p}_\chi - \vec{p}'_\chi)^2 + m_\phi^2} \langle f | \sum_j e^{i(\vec{p}_\chi - \vec{p}'_\chi)\cdot\vec{r}_{e,j}} | i \rangle
\end{eqnarray}
\end{widetext}
where $V$ is the normalization volume for the DM plane wave states.
By defining the momentum transfer $\vec{q} \equiv \vec{p}_\chi - \vec{p}'_\chi$, the matrix element becomes:
\begin{equation}
\langle f, \vec{p'}_\chi | \Delta H_{\chi e} | i, \vec{p}_\chi \rangle = \frac{g_e g_\chi}{V(q^2 + m_\phi^2)} \langle f | \sum_{j}e^{i\vec{q}\cdot\vec{r}_{e,j}} | i \rangle.
\end{equation}
Fermi's golden rule gives the scattering rate:
\begin{equation}
\Gamma = \int \frac{d^3 q}{(2\pi)^3} \sum_f \left| \langle f, \vec{p'}_{\chi} | \Delta H_{\chi e} | i, \vec{p}_\chi \rangle \right|^2 2\pi \delta(E_f - E_i - \omega).
\end{equation}
Using the interaction matrix element, we have
\begin{widetext}
\begin{eqnarray}
\Gamma &=& \int \frac{d^3 q}{(2\pi)^3} \frac{1}{V} \left( \frac{g_\chi g_e }{q^2 + m_\phi^2} \right)^2 \sum_f \left| \langle f | \sum_je^{i \vec{q} \cdot \vec{r}_{e,j}} | i \rangle \right |^2 2\pi \delta(E_f - E_i - \omega) \nonumber \\
&=& \int \frac{d^3 q}{(2\pi)^3} \left( \frac{g_\chi g_e}{q^2 + m_\phi^2} \right)^2 \left( \frac{q_0^2 + m_\phi^2}{q^2 + m_\phi^2} \right)^2 \frac{2\pi}{V} \sum_f \left| \langle f | \sum_je^{i \vec{q} \cdot \vec{r}_{e,j}} | i \rangle \right|^2 \delta(E_f - E_i - \omega)
\end{eqnarray}
\end{widetext}
We define the reference cross-section:
\begin{align}
{\sigma}_{\chi e} &= \frac{\mu_{\chi e}^2}{\pi} \left( \frac{g_\chi g_e}{q_0^2 + m_\phi^2} \right)^2
\end{align}
and define the DM form factor:
\begin{align}
F^{2}_{\mathrm{DM}}(q) &= \left( \frac{q_0^2 + m_\phi^2}{q^2 + m_\phi^2} \right)^2
\end{align}
where \( q_0 = \alpha m_e \).
The scattering rate can be written as
\begin{equation}
\Gamma = \int \frac{d^3 q}{(2\pi)^3} \, \frac{\pi}{\mu_{\chi e}^2} \, {\sigma}_{\chi e} \, F_{\text{DM}}^2(q) \, S(\vec{q}, \omega)
\end{equation}
where the structure factor is defined as: 
\begin{widetext}
\begin{eqnarray}
\label{equ: structure factor appendix}
    S(\vec{q},\omega)&=& \frac{2\pi}{V} \sum_f |\langle f |  \sum_ke^{i \vec{q} \cdot \vec{r}_k} | i \rangle|^2 \delta(E_f - E_i - \omega) \nonumber \\
   & &= \frac{2\pi}{V} \sum_f |\langle f | \sum_ke^{i \vec{q} \cdot \vec{r}_k} | i \rangle|^2 \frac{1}{2\pi} \int_{-\infty}^{\infty} dt e^{i \omega t} e^{-i (E_f - E_i) t} \nonumber \\
    &&= \frac{1}{V} \int_{-\infty}^{\infty} dt e^{i \omega t} \sum_f \langle f |\sum_k e^{i \vec{q} \cdot \vec{r}_k} | i \rangle \langle i |\sum_j e^{i H t}  e^{-i \vec{q} \cdot \vec{r}_j} e^{-i H t} | f \rangle \nonumber \\
    &&= \frac{1}{V} \int_{-\infty}^{\infty} dt e^{i \omega t} \sum_f \langle f | \sum_k e^{i \vec{q} \cdot \vec{r}_k(0)} | i \rangle \langle i |  \sum_j e^{-i \vec{q} \cdot \vec{r}_j(t)} | f \rangle \nonumber \\[10pt]
   & &= \frac{1}{V} \int_{-\infty}^{\infty} dt e^{i \omega t} \sum_f \langle f | \int d^3 r e^{i \vec{q} \cdot \vec{r}} \sum_k \delta^3 (\vec{r} - \vec{r}_k (0)) | i \rangle \langle i | \int d^3 r' e^{-i \vec{q} \cdot \vec{r}'} \sum_j \delta^3 (\vec{r}' - \vec{r}_j (t)) | f \rangle \nonumber \\
    &&= \frac{1}{V} \int_{-\infty}^{\infty} dt e^{i \omega t} \sum_f \langle f | \int d^3 r e^{i \vec{q} \cdot \vec{r}} \hat{n} (\vec{r}, 0) | i \rangle \langle i | \int d^3 r' e^{-i \vec{q} \cdot \vec{r}'} \hat{n} (\vec{r}', t) | f \rangle \nonumber \\
    &&= \frac{1}{V} \int_{-\infty}^{\infty} dt e^{i \omega t} \langle i | \hat{n} (\vec{q}, t) \hat{n} (-\vec{q}, 0) | i \rangle
\end{eqnarray}
\end{widetext}
where $\hat{n}(\boldsymbol{r},t)$ and $\hat{n}(\boldsymbol{q},t)$ are the electron density operators as defined in the main text.
The differential rate is
\begin{equation}
\frac{d\Gamma}{d\omega} = \int \frac{d^3 q}{(2\pi)^3} \, \frac{\pi}{\mu_{\chi e}^2} \, {\sigma}_{\chi e} \, F_{\text{DM}}^2(q) \, S(\vec{q}, \omega) \, \delta(\omega - \omega_\mathbf{q}) \,,
\end{equation}
where $\omega_\mathbf{q} =\textbf{q.v} - \frac{q^2}{2 m_\chi}$. The total differential rate per target mass is
\begin{equation}
\frac{dR}{d\omega} = \frac{\rho_\chi}{\rho_T m_\chi} \int d^3 v \, f(\vec{v}) \, \frac{d\Gamma}{d\omega}.
\end{equation}

Substituting for $\frac{d\Gamma}{d\omega}$, we obtain 
\begin{widetext}
\begin{align}
\frac{dR}{d\omega} &= \frac{\rho_\chi}{\rho_T m_\chi} \int \frac{d^3 q}{(2\pi)^3} \, \frac{\pi}{\mu_{\chi e}^2} \, {\sigma}_{\chi e} \, F_{\text{DM}}^2(q) \, S(\vec{q}, \omega) \int d^3 v \, f(\vec{v}) \, \delta(\omega - \omega_{\vec{q}}) \nonumber \\
&= \frac{\rho_\chi}{\rho_T m_\chi} \int \frac{d^3 q}{(2\pi)^3} \, \frac{\pi}{\mu_{\chi e}^2} \, {\sigma}_{\chi e} \, F_{\text{DM}}^2(q) \, S(\vec{q}, \omega) \, g(\vec{q}, \omega, t)
\end{align}
\end{widetext}

where \(g(\vec{q}, \omega, t)\) is the velocity integral:
$g(\vec{q}, \omega, t) = \int d^3 v \, f(\vec{v}) \, \delta(\omega - \omega_{\vec{q}})$.
We use these expressions in the main text.
\begin{figure*}[!htbp]
\centering
\includegraphics[width = 2.1\columnwidth]{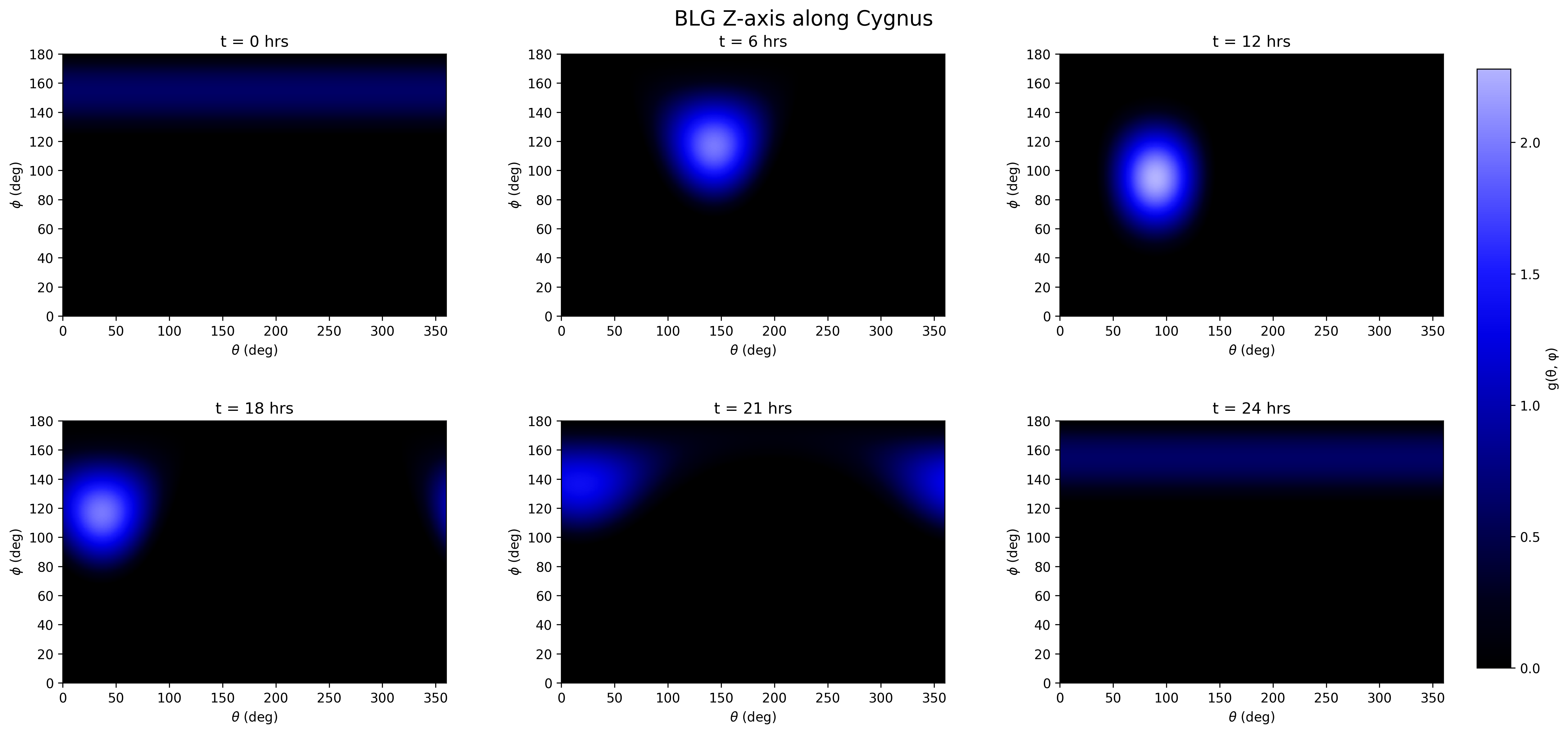}\\
\includegraphics[width =2.1\columnwidth]{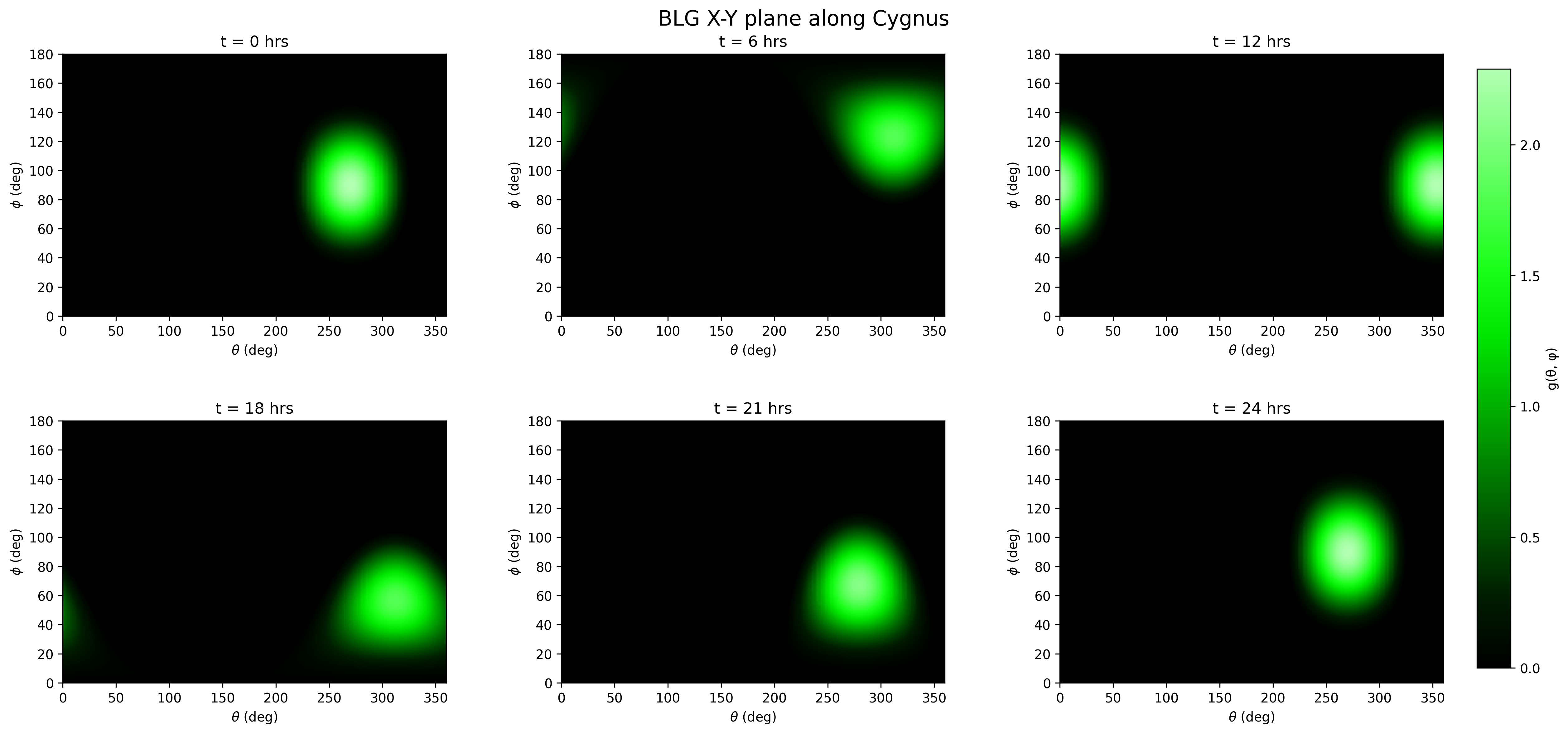}
\caption{ { The kinematic function $g(\vec{q}, \omega, t)$  for orientations 1 and 2 as function of angular coordinates ($\theta$ and $\phi$) with $m_\chi$ = 100 keV, $E_{th}$ = 100 meV, $q$ = 90 eV and $\omega$ = 0.15 eV,  plotted for various time intervals.} }
\label{fig: appendix Gfunction 1}
\end{figure*}

\begin{figure*}[ht]
\centering
\includegraphics[width=2.1\columnwidth]{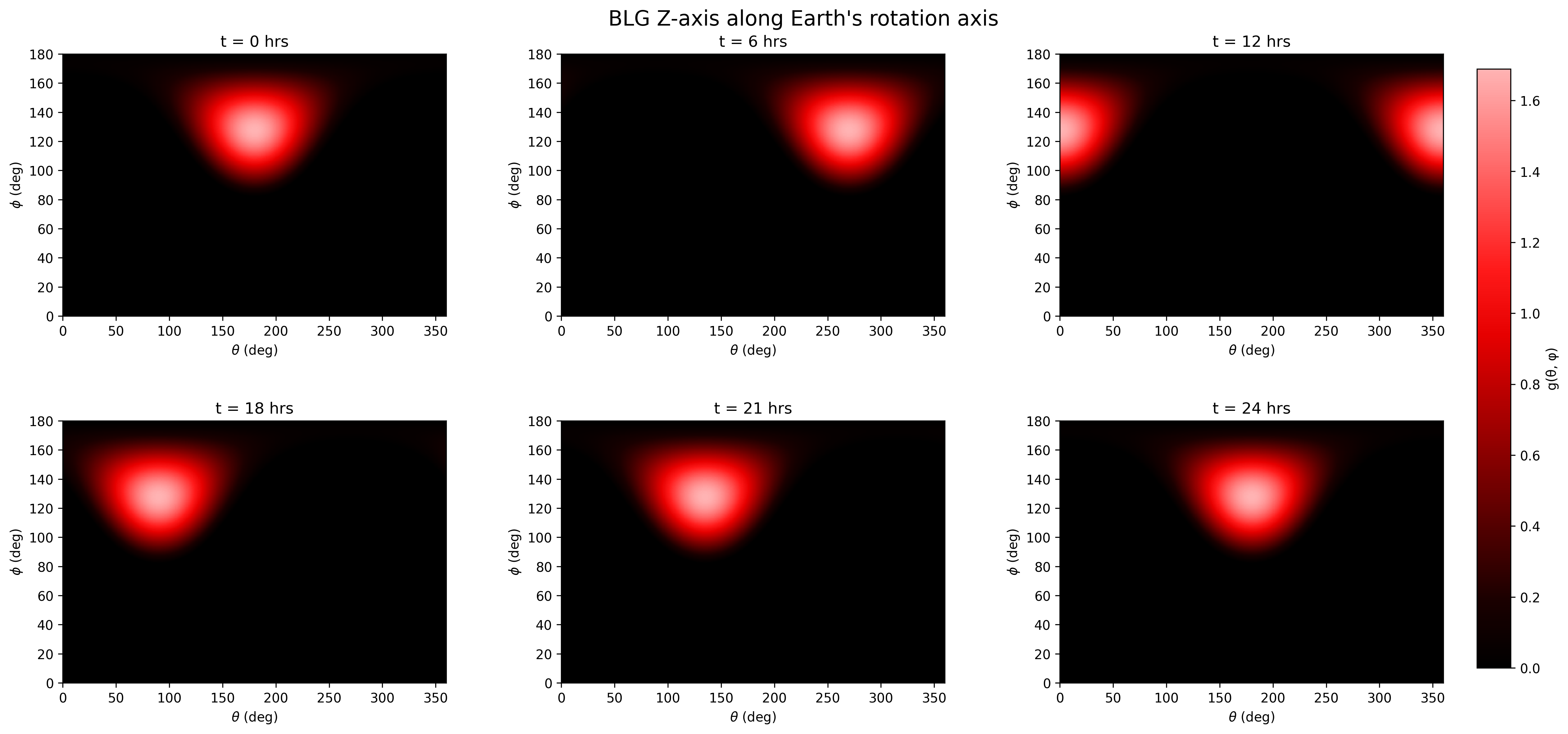}   \\
\includegraphics[width=0.7\columnwidth,height=5cm]{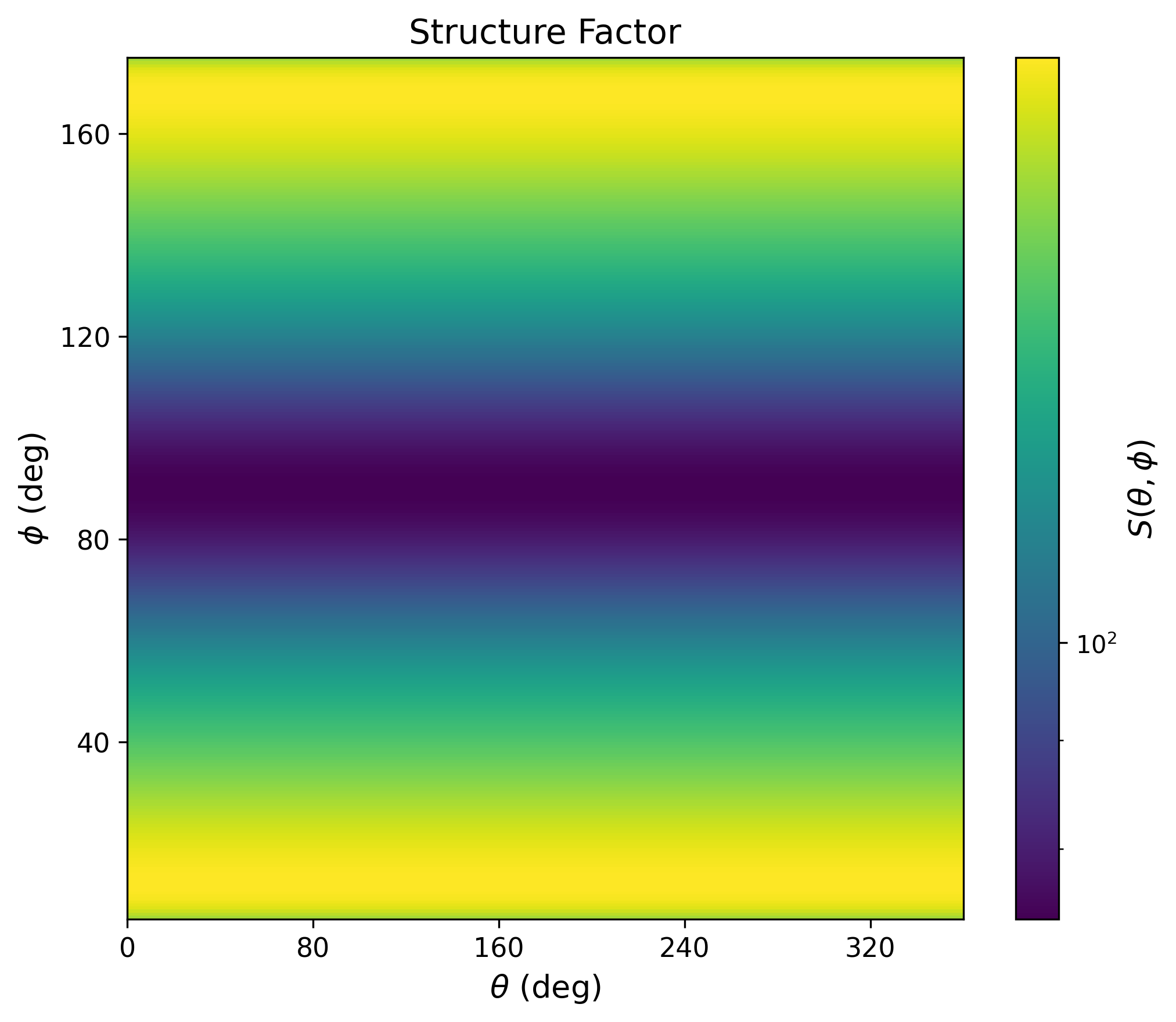}
\caption{{The kinematic function $g(\vec{q}, \omega, t)$ for orientation-3 and structure factor as functions of the angular coordinates ($\theta$ and $\phi$) with $m_\chi$ = 100 keV, $E_{\rm th}$ = 100 meV, $q$ = 90 eV and $\omega$ = 0.15 eV. The kinematic function is plotted for various time intervals. If bright regions of the kinematic function overlap with brighter regions of the structure factor, then we get a larger rate at that time.} }
 \label{fig: appendix Gfunction 2}
\end{figure*}
\section{Derivation of Velocity Integral}
\label{app: Derivation velocity integral}
The galactic frame DM velocity distribution is given by,
\begin{equation}
f_{\rm gal}(\mathbf{w}) = \frac{1}{N_0} 
\exp\left[-\frac{\mathbf{w}^2}{v_0^2}\right]\Theta(v_{\rm esc}-|\mathbf{w}|)
\end{equation}
where $\mathbf{w}$, $v_0$, $v_{\rm esc}$ are the galactic frame DM velocity, velocity dispersion, and escape velocity, respectively, and $N_0$ is the normalisation constant, same as in Eq.~\eqref{equ: G - function No normalization constant}. Now the DM velocity relative to the BLG coordinate frame is  $\textbf{v}=\textbf{w}-\textbf{v}_E$ or $\textbf{w}=\textbf{v}+\textbf{v}_E$, where $\mathbf{v}_E$ is the Earth (lab) velocity in the BLG coordinate system. As the velocity distribution is isotropic, substituting the expression for $\textbf{w}$ in $f_{\rm gal}(\mathbf{w})$ gives the DM velocity distribution in the BLG coordinate frame,i.e, $f(\mathbf{v})=f_{\rm gal}(\mathbf{v+v_E)}$. We want to evaluate the velocity integral,
\begin{widetext}
\begin{eqnarray}
\label{equ: G - function definition appendix}
g(\vec{q}, \omega, t) &=& \int d^3 v \, f(\vec{v}) \, \delta\left(\omega - \textbf{q.v} + \frac{q^2}{2 m_\chi}\right) \nonumber \\
 &=& \int \frac{d^3v}{N_0} \exp\left(-\frac{|\vec{v} + \vec{v}_E|^2}{v_0^2}\right) \delta\left(\omega - \vec{q} \cdot \vec{v} + \frac{q^2}{2m}\right)\, \Theta(v_{\rm esc} - |\vec{v} + \vec{v}_E|) \nonumber \\
& =& \int \frac{d^3u}{N_0} \, \exp\left(-\frac{|\vec{u}|^2}{v_0^2}\right) \delta\left(\omega - \vec{q} \cdot (\vec{u}-\textbf{v}_E) + \frac{q^2}{2m}\right)\, \Theta(v_{\rm esc} - |\vec{u}|)
\end{eqnarray}
\end{widetext}
where we have used the substitution $\vec{u} = \vec{v} + \vec{v}_E$ (or $\vec{v} = \vec{u} - \vec{v}_E$). The Jacobian of this transformation is unity and $d^3v = d^3u$. Aligning $\vec{q}$ along the z-axis, we use spherical velocity coordinates where 
$d^3u = u^2 \sin\theta \, du \, d\theta \, d\phi$ and 
$\vec{q} \cdot \vec{u} = qu \cos\theta$. The integral becomes:
\begin{widetext}
\begin{eqnarray}
 g(\vec{q}, \omega, t) &=& \int_0^{v_{\rm esc}} \int_0^{\pi} \int_0^{2\pi} 
\sin\theta \, d\phi \, d\theta \, du \,
\frac{u^2}{N_0} 
\exp\!\left(-\frac{u^2}{v_0^2}\right)  
\delta\!\left(\omega + \vec{q}\!\cdot\!\vec{v}_E
- q u \cos\theta + \frac{q^2}{2m}\right).
\end{eqnarray}
\end{widetext}
The $\phi$ integration gives a $2\pi$ term. For the $\theta$ integration, 
substitute $x = \cos\theta$, so $dx = -\sin\theta \, d\theta$.
The delta function constraint gives:
\begin{eqnarray}
\omega+\vec{q} \cdot \vec{v}_E - qux_0 + \frac{q^2}{2m} = 0 \nonumber \\
\end{eqnarray}
or
\begin{eqnarray}
\quad x_0 = \frac{\omega +\vec{q} \cdot \vec{v}_E+ q^2/2m}{qu} 
\end{eqnarray} is the root of the equation
and the integral can be written as
\begin{widetext}
\begin{eqnarray}
 g(\vec{q}, \omega, t)&= \dfrac{2\pi}{N_0} \displaystyle\int_0^{v_{\rm esc}} \displaystyle\int_{-1}^{1} \,dx\,du\, 
u^2 \exp\!\left(-\dfrac{u^2}{v_0^2}\right)\dfrac{
\delta\!\left(x-x_0\right)}{|qu|} 
\end{eqnarray}
\end{widetext}
The delta function integration contributes $1/|qu|$ when $|x| \leq 1$, requiring:
$\left|\dfrac{\omega + \vec{q} \cdot \vec{v}_E+q^2/2m}{qu}\right| \leq 1.$
Finally
\begin{widetext}
    \begin{eqnarray}
g(\vec{q}, \omega, t) &=& \frac{2\pi}{|q|N_0} \int_0^{v_{\rm esc}} du\, u\exp\left(-\frac{u^2}{v_0^2}\right)
 \Theta\left(1 - \left|\frac{\omega +\vec{q} \cdot \vec{v}_E+ q^2/2m}{qu}\right|\right)\nonumber \\
 &=& \frac{\pi v_0^2}{|q| N_0} \left[\exp\left(-\frac{v_{-}(q,t)^2}{v_0^2}\right) - \exp\left(-\frac{v_{\rm esc}^2}{v_0^2}\right)\right],
\end{eqnarray}
\end{widetext}
since the step function constraint translates to:
\begin{equation}
u \geq \frac{|\omega + \vec{q} \cdot \vec{v}_E+ q^2/2m|}{|q|},
\end{equation}
where
\begin{align}
v_{-}(q,t) &= \min \left\{ v_{\text{esc}}, \frac{\omega}{q} + \frac{q}{2 m_\chi} + \hat{q} \cdot \vec{v}_{\text{E}}(t) \right\}.
\end{align}

\section{Daily Modulation}
\label{app:daily_modulation}
As the scattering rate is the product of the kinematic function $g(\vec{q},\omega)$ and structure factor $S(\vec{q},\omega)$, we can also understand the daily modulation by looking at the time variation of these functions. In Fig.~\ref{fig: appendix Gfunction 1} and Fig.~\ref{fig: appendix Gfunction 2}, these are plotted as functions of angular coordinates ($\theta$ and $\phi$) of the $\mathbf{q}$ vector, with $m_\chi$ = 100 keV, $E_{\rm th}$ = 100 meV, $q$ = 90 eV and $\omega$ = 0.15 eV. When the brighter regions of the $g(\vec{q},\omega)$ overlap with the brighter regions of the $S(\vec{q},\omega)$, we get a larger rate during that time and a lower rate when they don't. In orientation-1, we see that the brighter regions of the $g(\vec{q},\omega)$ and structure function overlap initially, then the brighter region of the $g(\vec{q},\omega)$ moves to a less-bright region of the structure factor, and then the brighter region overlaps again, explaining the modulation. Instead, in orientation-2, the brighter regions don't overlap initially; around $t$ = 12 and 16 hours, there is maximum overlap; and then again at $t$ = 24 hours, there is minimum overlap. In orientation-3, there is no change in the $g(\vec{q},\omega)$ and structure function overlap region, and as a result, there is no modulation.

\bibliography{ref.bib}
\end{document}